\documentclass[iop]{emulateapj}
\usepackage{apjfonts}
\usepackage{graphicx}
\usepackage[breaklinks,colorlinks,urlcolor=blue,citecolor=blue,linkcolor=blue]{hyperref}
\journalinfo{The Astrophysical Journal, 2016, in press}

\shorttitle{Fading Coronal Structure and the Onset of Turbulence in the
Young Solar Wind}
\shortauthors{DeForest et al.}

\newcommand{\old}[1]{}
\newcommand{\new}[1]{#1}

\begin{document}

\title{Fading Coronal Structure and the Onset of Turbulence in the Young
Solar Wind}

\author{C. E. DeForest\altaffilmark{1},
W. H. Matthaeus\altaffilmark{2},
N. M. Viall\altaffilmark{3}, and
S. R. Cranmer\altaffilmark{4}}

\altaffiltext{1}{Southwest Research Institute, 1050 Walnut Street, Boulder,
CO, USA}
\altaffiltext{2}{University of Delaware}
\altaffiltext{3}{NASA / Goddard Space Flight Center}
\altaffiltext{4}{University of Colorado Boulder}

\begin{abstract}
Above the top of the solar corona, the young slow solar wind transitions
from low-$\beta$, magnetically structured flow dominated by radial
structures, to high-$\beta$, less structured flow dominated by hydrodynamics.
This transition\new{, long inferred via theory,} is readily apparent in the sky region close to {10\arcdeg}
from the Sun, in processed, background-subtracted solar wind images.
We present image sequences collected by the \emph{STEREO}/HI1 instrument
in 2008 Dec, covering apparent distances from approximately {4\arcdeg}
to {24\arcdeg} from the center of the Sun and spanning this transition
in large-scale morphology of the wind. We describe the observation
and novel techniques to extract evolving image structure from the
images, and \old{show}\new{we use those data and techniques to present and quantify the clear textural shift in the apparent structure
of the corona and solar wind in this altitude range.  We demonstrate} 
that the change in apparent texture is due both to anomalous fading
of the radial striae that characterize the corona, and to anomalous
relative brightening of locally dense puffs of solar wind that we
term ``flocculae;'' and show that these phenomena are \new{inconsistent
with smooth radial flow, but} consistent
with onset of hydrodynamic or MHD instabilities leading to a turbulent
cascade\old{, and inconsistent with smooth flow} in the young solar wind.
\end{abstract}


\section{Introduction}
\label{sec:Introduction}

The solar corona is largely structured by the magnetic field of the
Sun. At moderate solar altitudes above 2--5 $R_{\odot}$, the corona
is highly anisotropic, consisting primarily of radial and near-radial
structures such as rays (\citealt{Saito1965,NewkirkHarvey1968}),
streamers (\citealt{Bohlin1970}), pseudostreamers (\citet{WangEtal2007},
and similar dense, open radial structures (\citealt{AntiochosEtal2011}),
all of which may collectively be called ``striae'' when imaged remotely.
The structure is apparent in coronagraphic images of the K corona
(Thomson-scattered light): between about {1\arcdeg} and {10\arcdeg}
(4~$R_{\odot}$--40~$R_{\odot}$) from the Sun, the appearance
of the corona is dominated by radial striae that reflect density differences
between the various magnetically structured features. This structuring
\old{reflects}\new{arises from anisotropic} magnetic structure that is easily detected in the mid corona
with Thomson scattering (\citealt{MacQueenEtal1974}), in the EUV
corona via collisional line emission (\citealt{WalkerEtal1988}),
and even via dispersal of injected particles in the low corona (\citealt{RaymondEtal2014}).
In the mid corona, the magnetic field becomes largely radial (\citealt{Hundhausen1972}),
and variances in flow and density across the magnetic field give rise
to the radial striae. In addition to the striae, various localized
transient features may be seen: coronal mass ejections (CMEs) and
related ejecta (\citealt{GoslingEtal1974,McComasEtal1991}), bright
``Sheeley blobs'' (\citealt{SheeleyEtal1997}) and related small
dense ``Viall puffs'' that are identifiable in the solar wind (\citealt{ViallEtal2010,ViallVourlidas2015}),
and myriad small, hard-to-separate density fluctuations that can be
identified by motion-filter analysis (\citealt{DeForestEtal2014}).

The radial striae in coronal structure have been well observed for
decades. At some scales they appear to be preserved in the solar wind,
as the boundaries between the ``fast'' solar wind that is associated
with coronal holes (\citealt{Zirker1977}) and the ``slow'' solar
wind that is associated with the streamer belt and related structures.
These striated density structures are often interpreted to be outlining
and contrasting magnetic structures, i.e., flux tubes, in the sub-Alfv\'{e}nic
corona. In the corona, the plasma within these flux tubes is fully
interactive with regions above and below via \new{magnetohydrodynamic (}MHD\new{)} waves (\citealt{AlettiEtal2000}),
and in fact MHD waves are routinely observed directly via density
fluctuations and Doppler shifts in the low corona (\citealt{DeForestGurman1998,TomczykMcIntosh});
and adjacent flux tubes can also interact through magnetic reconnection
(e.g., \citealt{RappazzoEtal2012}). Both of these effects may be
viewed as consequences of the anisotropic turbulent MHD relaxation
that is expected in the presence of a strong guide field (\citealt{MontgomeryTurner1981,ShebalinMatthaeusMontgomery1983}),
especially at low plasma $\beta$ (\citealt{ZankMatthaeus92}). These
effects preserve structural inhomogeneities introduced by the connectivity
of adjacent flux tubes in the mid corona to different portions of
the solar surface. In this light, the anisotropic, striated appearance
of coronal flux tubes can be seen as \old{consequences}\new{a consequence} of differing rates
of parallel and cross-field relaxation and mixing \textbf{(}\citealt{CranmerEtal2015,OughtonEtal2015}).

In the inner heliosphere just outside the Alfv\'{e}n surface, where the
bulk speed first exceeds the fast mode magnetohydrodynamic (MHD) wave
speed, MHD signals cannot propagate back into the corona. However,
when sufficiently strong gradients are present, wave signals can still
generally overtake lateral expansion and produce local mixing and
a local turbulent cascade. Furthermore, with increasing $\beta$ in
the outer corona and inner heliosphere the lateral stabilization afforded
by the outer corona's radial magnetic field fades to insignificance,
enabling isotropization of the fluctuations in velocity and density.\footnote{Even for moderate magnetic field strengths, at sufficiently small
scales deep in the inertial range ($\ll1$~Mm) it is expected that
the anisotropizing effects of the magnetic field persist, because
the Alfv\'{e}n crossing time at length scale $\ell$ decreases as $\sim\ell$
while the nonlinear time scale decreases only as $\sim\ell^{2/3}$.
The present discussion refers to the outer scale and larger $\ell\gg$1~Mm. } Regions of high shear or high compression in the wind provide a theoretical
opportunity for rapid turbulent evolution over very large scales of
multiple Gm in the super-Alfv\'{e}nic wind due to hydrodynamic instabilities.
Stream interaction regions, which begin to develop in the inner heliosphere,
are a textbook example of such larger scale interaction (\citealt{Hundhausen1972});
and long-lived interactions form the familiar corotating interaction
regions (\citealt{GoslingPizzo1999}). High shear regions may be viewed
as regions where energy is more strongly injected into fluctuations,
harnessing free energy from differential speed of nearby streams.
An example of shear-driven turbulence at kinetic scales (\citealt{KarimabadiEtal2013})
illustrates the complexity of local structure that can be rapidly
generated from an initially smooth shear flow. This leads to the supposition
that hydrodynamic instabilities associated with differing flow speeds,
densities, and wave speeds in adjacent striae must lead to a turbulent
solar wind.

Turbulence is, in fact, routinely detected in the slow solar wind at 150 Gm
(1 AU)\new{, as summarized in a review by \citet{MatthaeusVelli2011}.  
The detected turbulence} \old{it} has an average (and highly variable) correlation length of
about 1 Gm, presumably increasing with distance \emph{r} as $r^{\alpha}$
with $0.5<\alpha<1$ closer to the Sun (\citealt{MatthaeusEtal2005,BreechEtal2008}).
\new{Comparisons of wind speed vs. temperature at different altitudes point to
ongoing turbulent processing between 0.3 and 4.5 AU (\citealt{ElliottEtal2012}).  
Further, } the idea that structures smaller than 1 Gm in the solar wind near
Earth are at least partially due to turbulent mixing was recently
corroborated using imaging of the fluctuating motion of comet tail
features (\citealt{DeForestEtal2015}). Such structures are ``small-scale''
from the imaging perspective, but are near the boundary between the
inertial and energy-containing scale ranges in solar wind turbulence
and are therefore ``large-scale'' from the turbulence theory perspective.

Not all Gm-scale structures in the solar wind arise from turbulence
above the corona. Coronal streamers appear to continuously emit trains
of small density puffs into the newly formed solar wind, with characteristic
radial size scales of a couple of solar radii and time scales of $\sim$90
minutes (\citealt{ViallVourlidas2015}); and we refer to these as
``Viall puffs.''
Per-event studies using \emph{STEREO}/COR and \emph{STEREO}/HI1
show that these quasi-periodic density structures flow with the slow
solar wind, accelerating from 90 km s$^{-1}$ at 2.5 $R_{\odot}$
to 285 km s$^{-1}$ at 50 $R_{\odot}$, and \new{that they} expand radially by approximately
the same factor in that range (\citealt{ViallEtal2010,ViallVourlidas2015}).
Viall puffs are too small to follow through the \emph{STEREO}/HI2
field of view, due to its lower spatial resolution and longer integration
times \textemdash{} but they apparently often survive to 1~AU, where they are
observed by in situ instruments and identified by composition and
periodicity (\citealt{ViallKepkoSpence2009,KepkoEtal2016}); and they
can drive dynamics in Earth's magnetosphere (\citealt{KepkoEtal2002,ViallSpenceKasper2009}).

The LASCO C-3 coronagraph (\citealt{BruecknerEtal1995}) flown on
\emph{SOHO} (\citealt{DomingoFleckPoland1995}), produces white-light
images out to {7.5\arcdeg} (30 $R_{\odot}$) from the Sun's apparent
location. It has been used to detect striae at all position angles,
including in the coronal holes, to the limits of its field of view
(e.g., \citealt{DeForestPlunkettAndrews2001}). More recently, the
SECCHI instrument suite (\citealt{Howard2008}) on board the \emph{STEREO}
mission has provided nearly continuous visible light imaging from
the inner corona through {90\arcdeg}, via the HI1 and HI-2 instruments
(\citealt{EylesEtal2009}). These instruments were primarily intended
to view CMEs (\citealt{HarrisonDavisEyles2005}) but with the advent
of deep-field background subtraction techniques (\citealt{DeForestEtal2011}),
they can be used to view both the disposition of the striae at the top
of the corona, and the corona's transition to the solar wind at the related
locations of the Alfv\'{e}n surface (at which the wind speed first exceeds
the speed of Alfv\'{e}n or field-aligned fast-mode waves) and the heliospheric
$\beta=1$ surface (at which the total gas pressure exceeds the magnetic
pressure).

The exact radial locations of the Alfv\'{e}n surface and the heliospheric
$\beta=1$ surface are not well constrained observationally.
\citet{DeForestEtal2014}
used \emph{STEREO}/COR2 observations of inbound fluctuations along
striae and in the coronal holes as evidence for the Alfv\'{e}n surface
being at least 12~$R_{\odot}$ from the Sun over polar coronal holes
and 15~$R_{\odot}$ in the streamer belt. Models of coronal heating
and wind acceleration along open magnetic field lines also make predictions
of these locations. One-fluid ZEPHYR models (Cranmer et al. \citeyear{CranmerEtal2007,CranmerEtal2013})
reproduced the general latitudinal structure of the coronal wind at
solar minimum, as well as in-ecliptic fluctuations associated with
quiet Sun regions, and contained a modeled Alfv\'{e}n surface typically
between 7 and 15 $R_{\odot}$ and a $\beta=1$ surface typically between
20 and 50 $R_{\odot}$. These values are in general agreement with
the output of several independent three-dimensional heliospheric simulations
\citep[e.g.,][]{LionelloEtal2014,Cohen2015,FengEtal2015}, and place
these important surfaces somewhere in the lower half of the
\emph{STEREO}/HI1 field of view.

The HI1 field of view is square on the focal plane, and it extends
from approximately {3.75\arcdeg} to {24\arcdeg}
($\sim$15~$R_{\odot}$ to 96~$R_{\odot}$)
from the Sun, though that entire range
is not available along any one solar-radial chord through the field
of view. An important textural transition can be seen in background-subtracted
images from that instrument: in the lower/inner part of the field
of view, at solar elongation angles near {5\arcdeg}, the radial
striae from the streamer belt and related structures dominate the
coronal plasma; while in the upper/outer part of the field of view,
at elongation angles near {20\arcdeg}, the radial striae no longer
dominate and the plasma takes a puffy, or flocculated, appearance.
This transition has, to our knowledge, not yet been described in the
literature despite nearly a decade of \emph{STEREO} operations. That
is perhaps because of the difficulty of separating the textural change
from other observationally important transitions occurring across this
field of view. In particular, between {4\arcdeg} and {20\arcdeg}
solar elongation, the Thomson scattering signal drops by a factor
of roughly 125, falling from roughly 10$\times$ to under 0.1$\times$
the average surface brightness (radiance) of the background starfield.
Thus, careful treatment of the background is essential to understanding
the shift in appearance.

In the present work, we investigate, characterize, and speculate on
the causes of the textural shift in the transition from coronal structure
to solar wind structure in the outer reaches of the solar corona,
across the HI1 field of view. In Section \ref{sec:Data-Methods}
we describe the data and our methods for preparing them; in Section
\ref{sec:Results} we present results of visual and quantitative analysis
of individual features (\ref{sub:Feature-propagation}) and the statistical
ensemble of the images (\ref{sub:Autocorrelation});
in Section \ref{sec:Discussion}
we discuss possible explanations for the transition in visual texture
from the outer corona to the young solar wind; and in
Section \ref{sec:Conclusions}
we conclude and summarize the results and needed future work.

\section{Data \& Methods}
\label{sec:Data-Methods}

We used synoptic images from the HI1 imager (\citealt{EylesEtal2009})
in the SECCHI suite (\citealt{Howard2008}) aboard the \emph{STEREO}-A
spacecraft, over a single 15 day interval: 2008 Dec 15 to 2008 Dec
29 inclusive (a period near solar minimum, with well defined streamer belts).
The principal concern with imaging wind evolution in this regime is
the intrinsic dependence of the solar wind radiance on distance $r$
from the Sun: as the optically thin solar wind propagates outward
at roughly constant speed, its density falls as $r^{-2}$ -- and its
illumination function also falls as $r^{-2}$. These factors combine
with the inherent $r$ length scale of of the line of sight, to yield
an $r^{-3}$ dependence of \old{volumetric scattered emissivity}\new{apparent surface brightness}. That is
reflected in an intrinsic variation in the total Thomson scattered
radiance of $\varepsilon^{-3}$ at solar elongation angles $\varepsilon$
up to about {30\arcdeg}. The linear relationship between solar
distances and elongation angles, at $\varepsilon$ values below
{30\arcdeg},
and its breakdown at higher values, have been well described in the
literature (e.g., \citealt{HowardTappin2009}). The
radiance of the solar wind thus rapidly drops below the radiance of
the background starfield. It is important to note that $r$ is a
three-dimensional quantity, and when considering variations along
a line of sight, the impact parameter $b$ of a given line of sight
corresponds more directly to $\varepsilon$. This is discussed in
some detail in the Appendix, and illustrated there in Figure \ref{fig:Perspective-diagram-shows}. 

\begin{figure*}
\epsscale{1.11}
\plotone{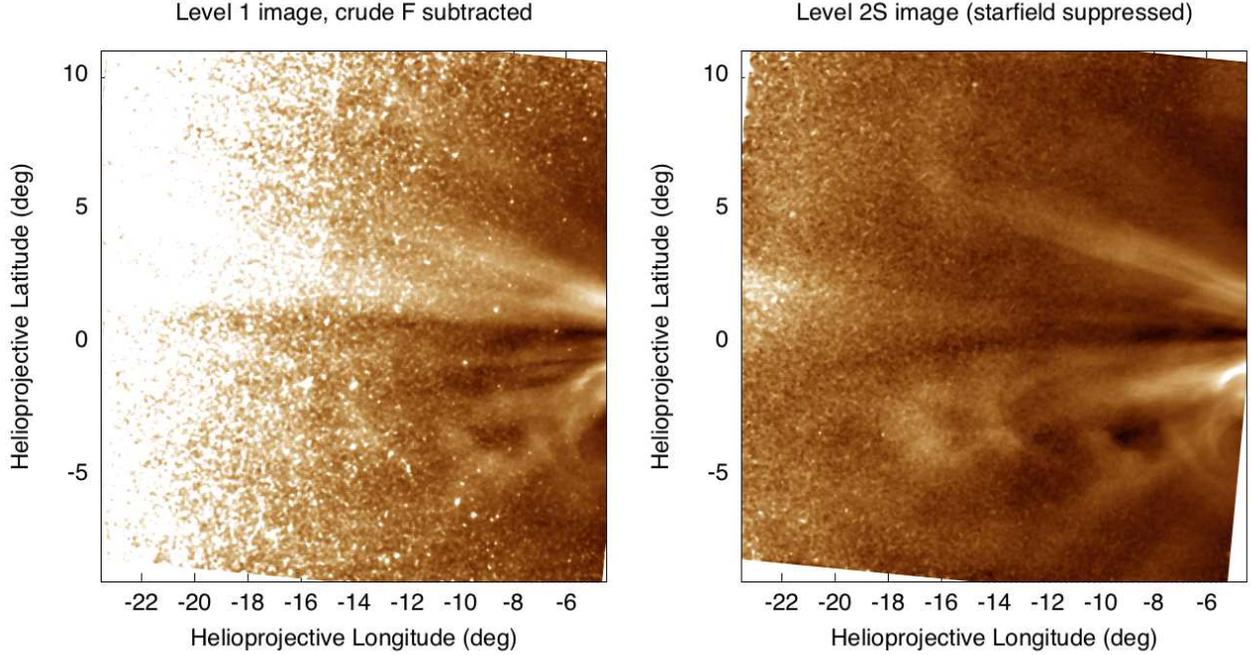}
\caption{\label{fig:HI1-example}Two copies of the same HI1A image, acquired
on 2008 Dec 16 20:09, show how the images are affected by starfield
subtraction processing. Both images have been radial-filtered by $\varepsilon^{3}$
scaling, to compensate for the secular falloff in intensity far from
the Sun. {\em LEFT:}
A direct HI1 image, with simple F coronal model subtracted;
{\em RIGHT:} L2S image suppresses the starfield by a factor of 30--100.}
\end{figure*}

Fortunately, the HI1 detector response to incident light is quite
linear, and the stars are nearly constant brightness. This enables
reprocessing of HI1 data to remove the fixed starfield, as described
by \citet{DeForestEtal2011}. Background-subtracted data produced
with their algorithm are available from the \emph{STEREO} Science Center,
as processing levels ``L2S'' and ``L2M''. The L2S data are subjected
to starfield subtraction using a local-brightness technique that attenuates
the background starfield by a factor of 30--100. The L2M data are further
processed with a Fourier-domain motion filter, which suppresses the
residual starfield by another factor of 10--30 but also suppresses
the quasi-stationary radial structures of interest to this study,
so we used the slightly noisier L2S data. The effect of L2S processing
is illustrated in Figure \ref{fig:HI1-example}.

We transformed the L2S images to conformal radial coordinates. These
are radial coordinates with a logarithmic scale in the radial direction,
which (with appropriate scaling of the azimuth and log-radius axes)
yields a conformal transformation: one that preserves shape, though
not orientation or scale, of small features (e.g., \citealt{Rudin1987,DeForestEtal1997}).
To avoid aliasing of the starfield and to reduce photon noise by pixel
averaging where possible, we resampled the images using optimized
spatial filtering keyed to the Jacobian derivative of the transformation
(\citealt{DeForest2004}). The conformal radial coordinate system
is particularly useful for two reasons: (1) it enables smoothing across
larger larger patches of instrument pixels in the outer reaches, to
further beat down background noise where necessary (as demonstrated
by \citealt{DeForestPlunkettAndrews2001}); and (2) it eliminates
the visually dominant lateral expansion of the plasma flow across
the HI1 field of view, and thereby simplifies visualization of both
the usual radial flow and deviations from it.

To further suppress the starfield, we median-filtered each image again,
across square patches subtending {0.6\arcdeg} in solar azimuth
in the polar coordinate system. This patch size is equivalent to $21\times21$
L1 pixels at the outer edge of the field of view, and $4\times4$
L1 pixels at the inner edge. 

Next, we removed broad brightness structures via unsharp masking with
the \emph{minsmooth }operator over a circular kernel (\citealt{DeForestHoward2015}).
Unsharp masking is the technique of highlighting small scales by subtracting
a smoothed version of an image, from the image itself. \emph{Minsmooth}
replaces broad-kernel convolution to produce the smooth ``background.''
It collects a low-percentile value of each pixel's neighborhood, further
blurring this neighborhood minimum map with broad-kernel convolution.
The result is an effective smooth background for the image, based
on feature spatial scale alone. Unsharp masked images made by subtracting
a \emph{minsmooth}ed copy of an original are approximately positive-definite,
although they contain only features on spatial scales smaller than
the applied neighborhood size. We applied \emph{minsmooth} unsharp
masking on two scales: {14\arcdeg} diameter to reduce diffuse
brightness in the images; and {4.2\arcdeg} diameter to highlight
marker structures in the outer corona and solar wind. 

Finally, to suppress residual noise, we averaged across three frames
in time. To prevent motion blur from the typical {1\arcdeg} of
outward motion in this interval, we shifted each triplet of frames
to their central time, based on a typical radial speed of 350 km~s$^{-1}$.
This reduced radial motion blur to under {0.2\arcdeg}. We selected
this speed by measuring outflow rates of several small features, assuming
they were close to the Thomson surface (i.e., that $b\approx r$ for
each feature). That assumption is not necessarily warranted for three
dimensional studies, but sufficed to identify the bulk motion of the
features and reduce motion blur. The measured projected speeds were
all within 300--400 km~s$^{-1}$, consistent with accepted values
for the slow solar wind speed.

\begin{figure}
\epsscale{0.98}
\plotone{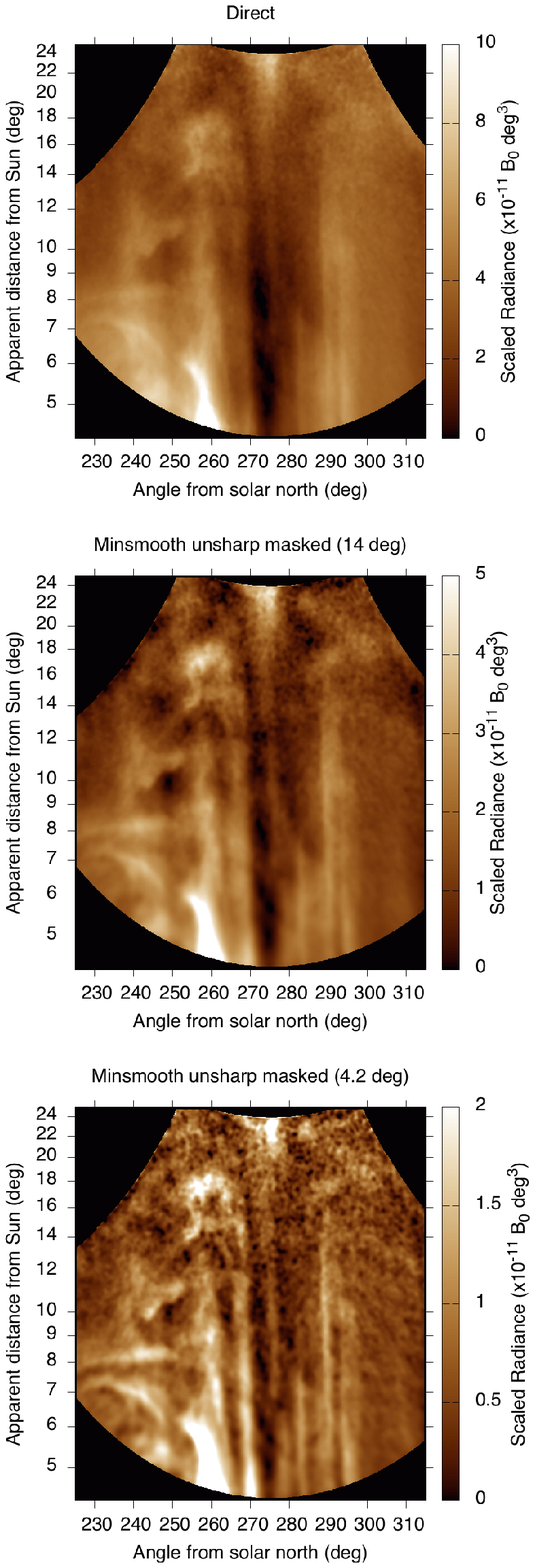}
\caption{\label{fig:conformal-radial}Radial structure is visible in conformal
radial coordinates. {\em TOP:}
the right frame from Fig. \ref{fig:HI1-example},
after further median filtering, shows fading radial structure.
{\em MIDDLE:}
minsmooth unsharp masking with a {14\arcdeg} wide kernel hides
diffuse brightness, revealing structure. {\em BOTTOM:} minsmooth unsharp
masking with a {4.2\arcdeg} wide kernel highlights both radial
structure and transient ejecta. See also the accompanying movie in
the digital version of this article, which depicts the full data
set as processed in the middle panel. For further analysis we used
frames similar to the middle panel.}
\end{figure}

Figure \ref{fig:conformal-radial} shows an example transformed frame
and the effect of \emph{minsmooth} unsharp masking on it. The top
panel is a direct copy of the right-hand panel of Figure \ref{fig:HI1-example},
after local median smoothing to remove the residual starfield. The
middle panel shows the effect of a {14\arcdeg} wide \emph{minsmooth}
unsharp mask, which removes broad diffuse structure without affecting
the coronal features. For the remainder of this work, we used these
{14\arcdeg} unsharp-masked images. The bottom panel shows the
effect of more aggressive \emph{minsmooth} unsharp masking, which
reveals small-scale features more clearly but also perturbs the image
on the spatial scales of interest for our textural study.

The unsharp masked images show radial structure, and its apparent
loss, quite strikingly: although the bottom portion of each image
is dominated by bright radial structures, and the $\varepsilon^{3}$
scaling should preserve apparent brightness (image pixel value) of
wind features as they propagate, the radial structures fade with altitude,
nearly vanishing somewhere between {12\arcdeg}--{18\arcdeg}
from the Sun. In this range, a more nearly isotropic pattern of small
puffs comes to dominate, giving the image a flocculated appearance
(so called because it resembles the texture of a sheep's side, or
that of a flocculated chemical solution, such as ale wort after the
cold break; \citealt{Papazian2003}). The flocculated texture is more
apparent in the more aggressively unsharp masked bottom panel of Figure
\ref{fig:conformal-radial}, which is included for reference only:
all further analysis used the {14\arcdeg} unsharp masked data.

\section{Results}
\label{sec:Results}

The primary result is clearly visible in the bottom two panels of
Figure \ref{fig:conformal-radial}: while the radial, striated structure
of the coronal streamer belt is obvious and dominates the images at
the lower reaches of the HI1 field of view (out to roughly {10\arcdeg},
or $\sim$40~$R_{\odot}$), that structure largely disappears in the
outer half of the field of view. That is surprising, because the $\varepsilon^{3}$
factor in the scaled brightness compensates for both feature expansion
and the decrease of the solar illumination function: features evolving
only with the average characteristics of the solar wind should maintain
the same apparent scaled brightness as they propagate through the
scaled images' field of view. We further tested and refined this result in two ways:
analysis of individual features and extraction and analysis of an image structure
function.

\subsection{Feature Propagation}
\label{sub:Feature-propagation}

\begin{figure*}
\epsscale{1.07}
\plotone{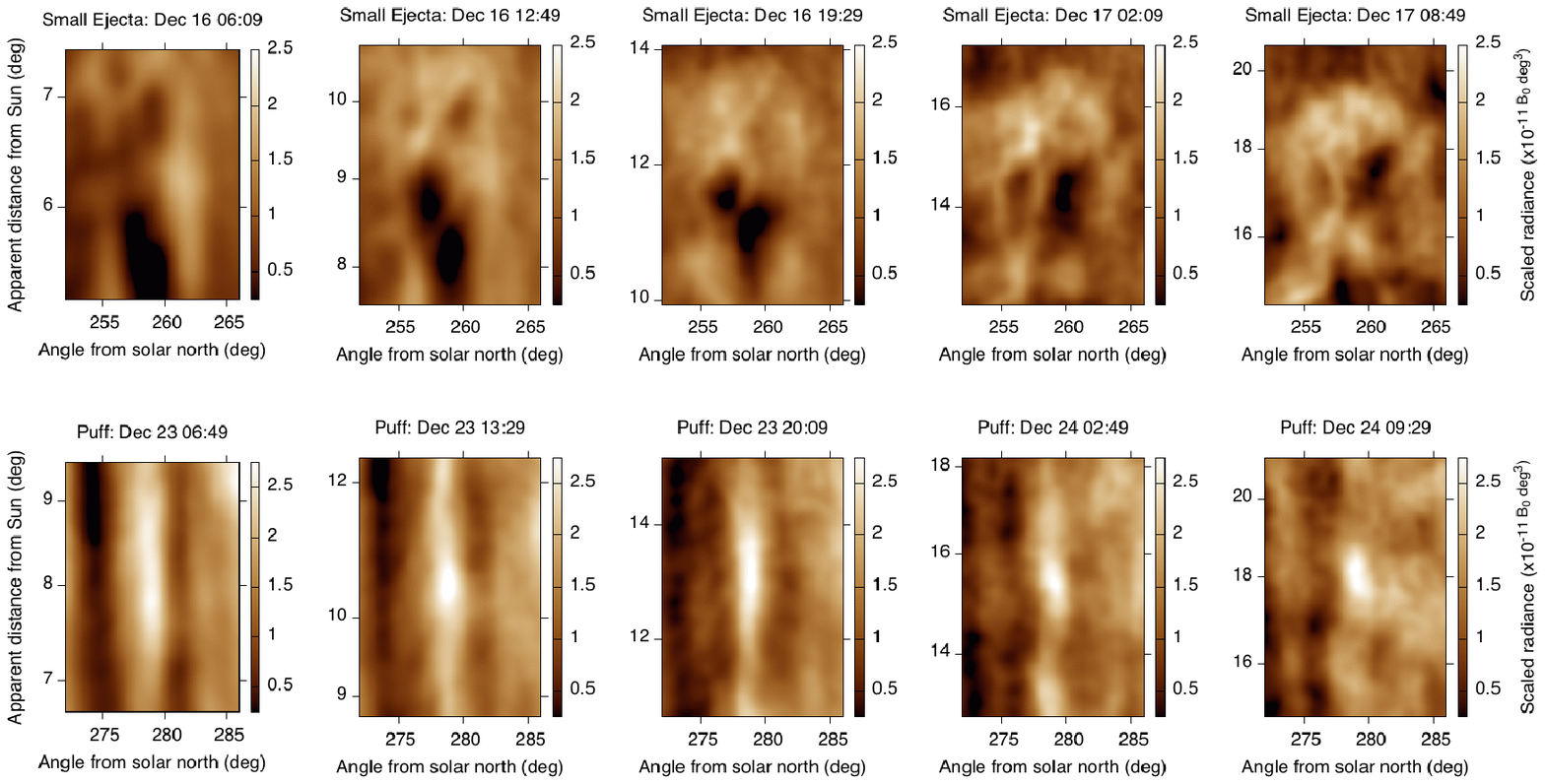}
\caption{\label{fig:Tracked-evolution-sequences}Tracked evolution sequences
of two typical bright features (identified inside of {10\arcdeg}
from the Sun) reveal that Equation \ref{eq:constant} holds: both
features maintain approximately constant brightness in the radially-scaled
images, as can be seen by reference to the color bars at right of each
panel. They also exhibit the well known ``pancake effect'' from
radial propagation; this appears as radial squashing in the conformal
polar coordinates. See also the corresponding movies in the digital
version of this article. {\em TOP:}
a mini-CME. {\em BOTTOM:} a localized, dense
puff within a stria.}
\end{figure*}

Radial features exhibiting the ``typical'' solar wind behavior of
constant-speed flow and conservation of mass are expected to maintain
\new{approximately} constant brightness in Figure \ref{fig:conformal-radial}. To detail
the above introductory discussion of expected brightness scaling,
the radiance (surface brightness $B$) of a feature is just
\begin{equation}
B_{feat}=\pi R_{\odot}^{2}B_{\odot}\sigma_{t}\int\left(n_{e}\frac{1+(\cos\chi)^{2}}{R^{2}}\right)ds,\label{eq:radiance}
\end{equation}
where $R_{\odot}$ is the solar radius, $B_{\odot}$ is the mean solar
radiance, $\sigma_{t}$ is the Thomson scattering cross section, $n_{e}$
is the local electron density, $\chi$ is scattering angle, $R$ is
distance from the scattering site to the Sun, and $s$ is distance
along a given line of sight (e.g., \citealt{HowardDeForest2012}).
From conservation of mass in a constant-radial-flow solar wind, we
observe that $n_{e}$ can be written 
\begin{equation}
n_{e}=n_{e0}R_{0}^{2}R^{-2}\label{eq:density-scaling}
\end{equation}
for a constant (against radius) equivalent density at the photosphere
$n_{e0}$ and a reference distance $R_{0}$ (e.g., 1~$R_{\odot})$.
Thus, Equation \ref{eq:radiance} simplifies in the constant-flow
case to
\begin{equation}
B_{simple}=\pi R_{\odot}^{2}B_{\odot}\sigma_{t}n_{e,0}R_{0}^{2}\int\left(1+(\cos\chi)^{2}\right)R^{-4}ds=k\,b^{-3},\label{eq:scaling}
\end{equation}
where $k$ is a constant of proportionality that includes all the
terms to the left of the integral and an additional geometrical factor,
and $b$ is the impact parameter of the line of sight relative to
the Sun. Since the angles under consideration in the HI1 images are
less than {30\arcdeg}, the small angle formula applies
and the relation $b\approx\varepsilon R_{obs}$ holds (where $\varepsilon$
is the observed separation angle between an image pixel and Sun center,
and $R_{obs}$ is the distance from the observer to the Sun). Thus,
for simple flow, 
\begin{equation}
B_{simple}\,\varepsilon^{3}\approx kR_{obs}^{-3},\label{eq:constant}
\end{equation}
which is constant against $\varepsilon$. We tracked several bright
features to show that radiance is preserved as expected; two examples
are given in Figure \ref{fig:Tracked-evolution-sequences}. The top
of row of panels shows a visually striking, compact CME that launched
on 2008 Dec 16. The bottom row shows a single bright puff embedded
inside a stria. Both features maintain approximately constant brightness
in the radially-scaled images.

\begin{figure*}
\epsscale{1.07}
\plotone{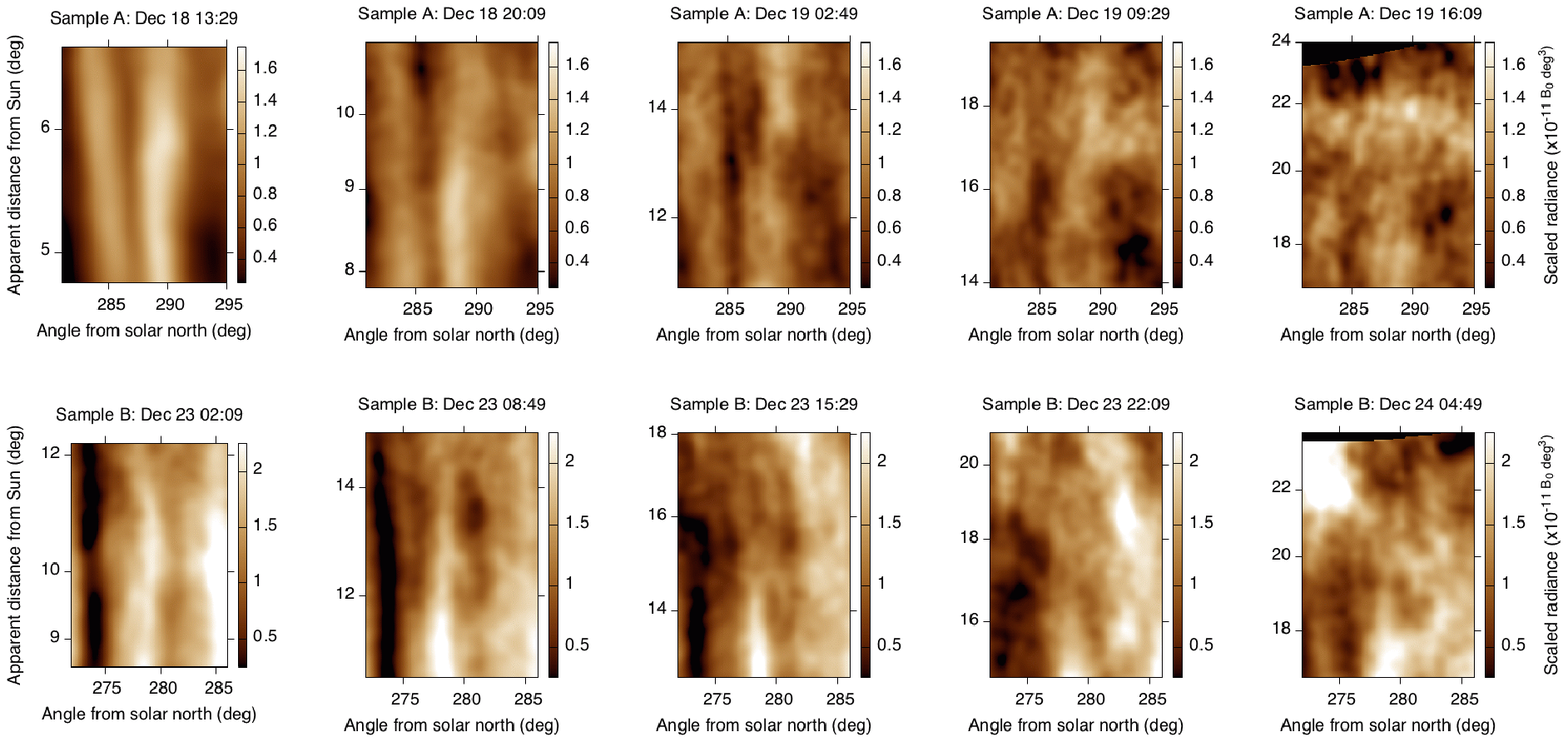}
\caption{\label{fig:Evolution-sequences-within}Evolution sequences within
two ``quiet'' striae (radial structures) show fading with distance.
The sequences were extracted as in Figure \ref{fig:Tracked-evolution-sequences},
but in locations with no clear puff. See also the corresponding movies in the
digital version of this article.}
\end{figure*}

By contrast, individual radial striae (at locations where there is
no particular localized feature to track) fade faster than $\varepsilon^{-3}$.
The right hand side of Figure \ref{fig:conformal-radial} shows this
effect in broad context: although the figure uses scaled brightness,
the striae visibly fade before reaching the outer limit of the field
of view. Figure \ref{fig:Evolution-sequences-within} shows two particular
examples of fading radial striae, extracted as in
Figure \ref{fig:Tracked-evolution-sequences}
to follow evolution of a particular patch of solar wind as it propagates
outward.

\begin{figure*}
\epsscale{1.05}
\plotone{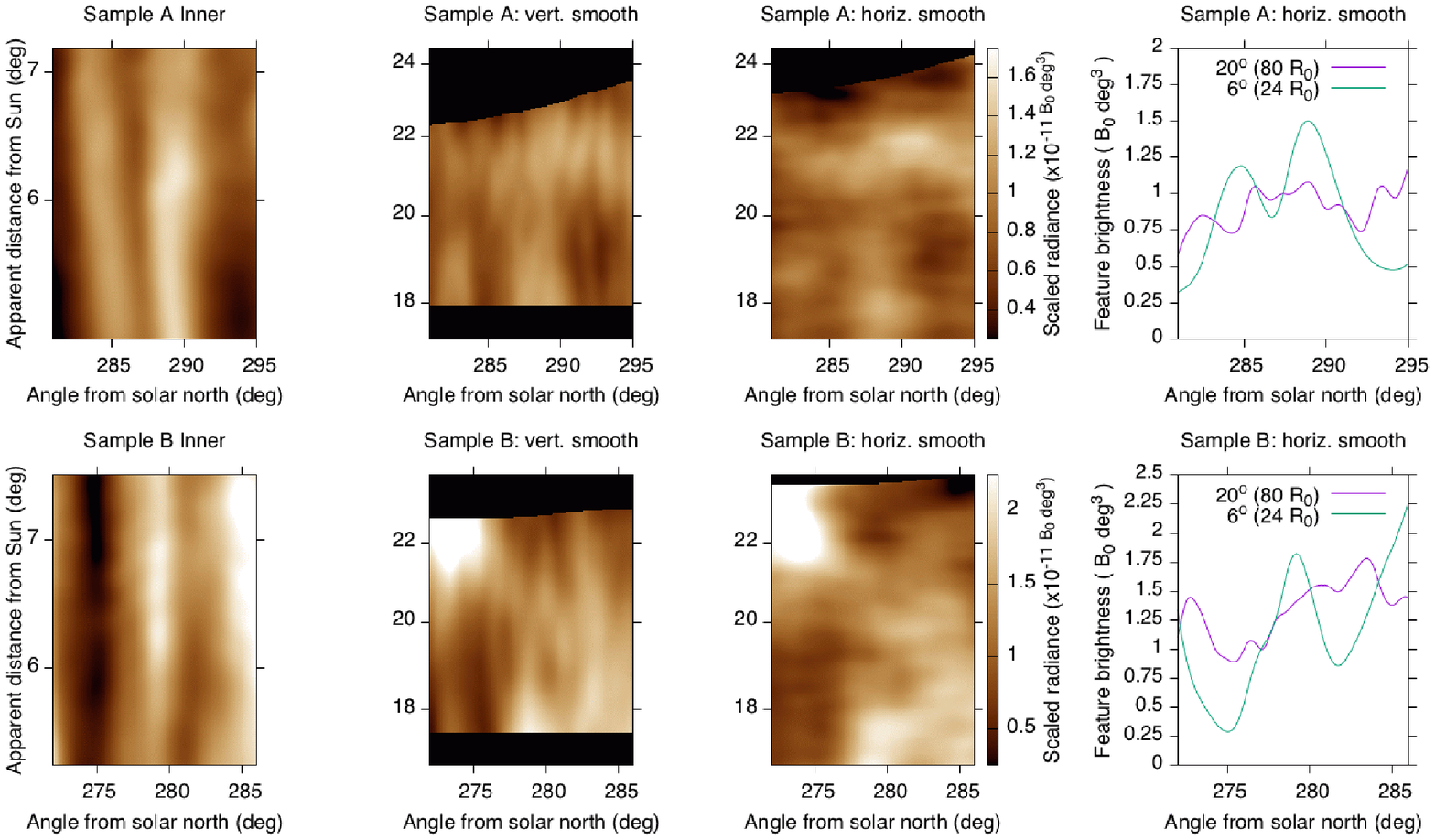}
\caption{\label{fig:Smoothing-the-final}Smoothing the final panels of Figure
\ref{fig:Tracked-evolution-sequences} reveals remnant striations
even beyond {20\arcdeg} (80~$R_{\odot}$). {\em LEFT:} The striations
at low altitude, from Figure \ref{fig:Evolution-sequences-within}.
{\em MIDDLE:} both vertical and horizontal smoothing reveal continued
vertical/radial structure
below the noise floor. {\em RIGHT:} plots of the structure amplitude show
3$\times$-5$\times$ attenuation of the striations' measured normalized
brightness between {6\arcdeg} (24~$R_{\odot}$) and {20\arcdeg}
(80~$R_{\odot}$). Uncertainty may be estimated from the residual
high spatial frequencies, as approximately $\pm$0.15~$B_{\odot}$~deg$^{-3}$
in each {1\arcdeg} wide spatial interval.}
\end{figure*}

Qualitatively, the striae (vertical stripes) in Figure \ref{fig:Evolution-sequences-within}
fade continuously from bottom to top in these tracked sequences, dropping
below the noise floor. To quantify this fading, and verify that it
is not an artifact of the rising noise floor, we smoothed the final
images in the vertical direction to bring down the noise floor still
further and identify whether the radial stria is still present. To
verify whether any identified structure was due to the smoothing direction,
we also carried out the identical smoothing operation in the horizontal
direction. The results are in Figure \ref{fig:Smoothing-the-final},
which shows clear remnant vertical structure at the smoothed noise
floor after smoothing across {1.5\arcdeg} of radial extent or
the equivalent ({6\arcdeg}) in azimuthal extent (all at {20\arcdeg}
elongation).

Although (just) visible in the smoothed images, the remnant radial
structure is attenuated by a factor of 3--5 at {20\arcdeg} elongation,
compared to its scaled brightness at {6\arcdeg}. Note that the
important parameter in Figure \ref{fig:Smoothing-the-final} is the
peak-to-peak amplitude of the visible striations, not their absolute
brightness: the necessary background subtraction and minsmooth unsharp
masking steps invalidate the zero-point measurement.

The fading of the radial striae indicates a real attenuation of their
emission compared to a radially propagating, mass-conserving, solar
wind whose speed is independent of radius. The result is strong: by
starting with the flat-fielded Level 1 data, avoiding any motion filtering
beyond simple starfield removal, performing a control analysis by
tracking compact features, and demonstrating persistence of the features
despite anisotropic smoothing, we have eliminated most credible causes
of potential confusion or loss-of-signal artifacts.

\begin{figure*}
\epsscale{1.05}
\plotone{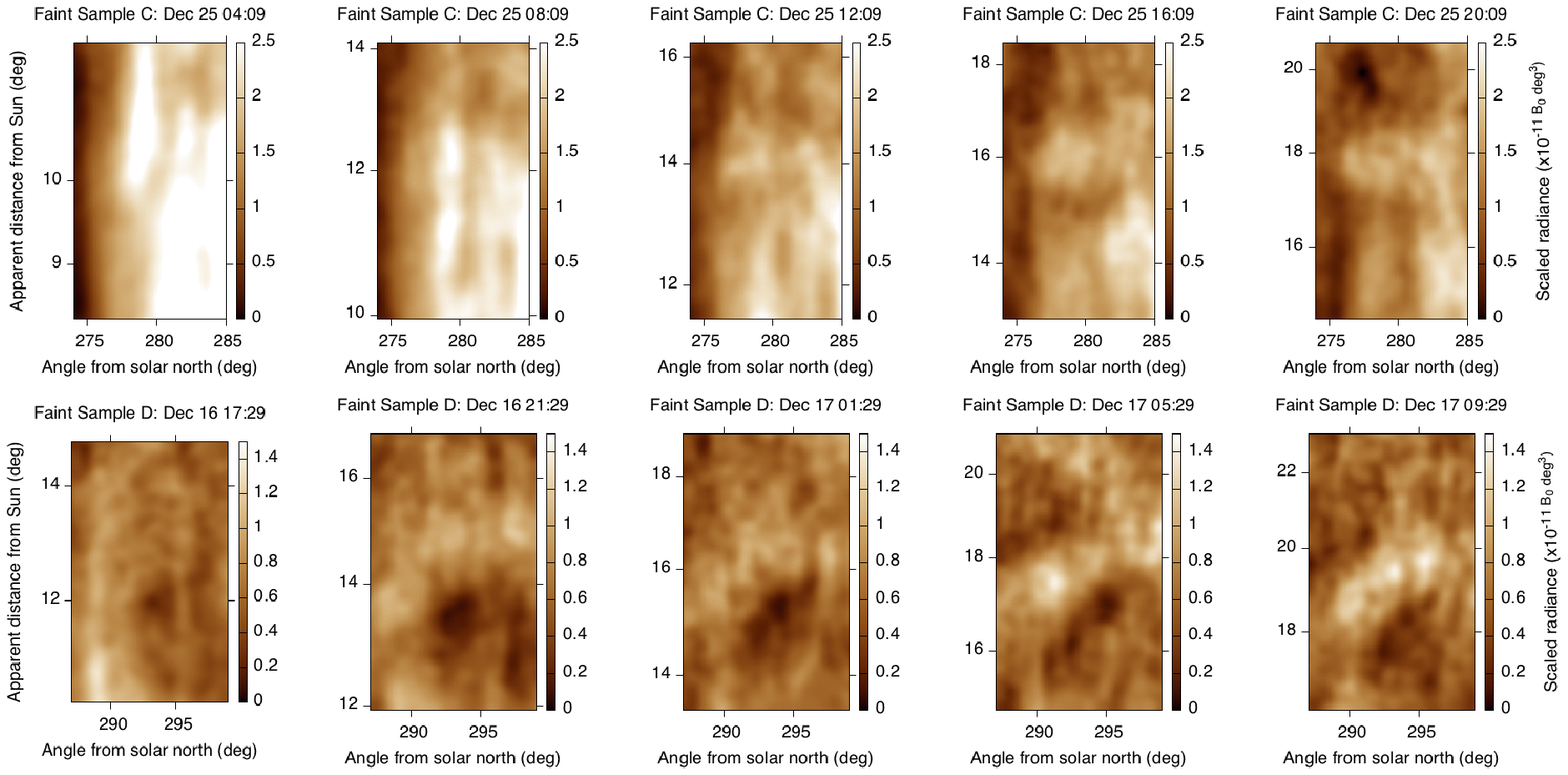}
\caption{\label{fig:Faint-relatively-compact}Faint relatively compact features
(``flocculae'') grow by comparison to the coronal striae, to generate
the flocculated appearance of the outer portion of the HI1 field of
view. These faint features (found outside of {15\arcdeg} from
the Sun) appear to ``fade in'' as they propagate across the field
of view. Each sequence shows, from left to right, a sequentially brightening
localized image feature subtending approximately {5\arcdeg}--{9\arcdeg}
of azimuth and roughly vertically centered in the image panel. {\em TOP:}
a floccule fades in between elongations of {10\arcdeg} to {18\arcdeg},
even as its host stria fades out. {\em BOTTOM:}
a floccule fades in between
elongations of {13\arcdeg}--{19\arcdeg}, in isolation from strong
striae.}
\end{figure*}

In parallel with the radial fading of the long, anisotropic radial
striae in the corona, smaller bright features become more prominent,
giving the plasma a puffy or flocculated appearance. The \citet{SheeleyEtal1997}
blobs and the \citet{ViallEtal2010} puffs arise from the low corona
and do not fade in scaled brightness as they propagate, becoming part
of the flocculated gestalt. In addition, there are small puffs with similar
size scales that fade in from invisibility in the moving frame of
reference of the wind itself, and are therefore visible primarily
in \old{teh}\new{the} outer field of view. We refer to these brightening puffs as
``flocculae.'' Figure \ref{fig:Faint-relatively-compact} illustrates
two of them. The flocculae are in general fainter and larger than
the puffs visible in the lower portion of the field of view, appear
to grow in scaled brightness as they propagate, and can best be detected
only outside of $\sim${10\arcdeg}, fading in even as
the striae fade out with altitude.

It is this combination of the fading striae and the growing flocculae
that gives rise to the flocculated appearance of the outer HI1 field
of view in properly background-subtracted image sequences. In particular,
if the HI1 image quality were degrading sufficiently to obscure the
striae, the flocculae and other localized features would be equally
obscured. The fact that particular classes of feature, propagating
with the solar wind, grow in amplitude, even as others fade, reveals
that simple loss of contrast or related image degradation is not causing
the change in image texture across the field of view: the textural
shift reflects a shift in the spatial form of density fluctuations
across this field of view.

\subsection{Structure Function Analysis}
\label{sub:Autocorrelation}

In addition to anecdotal analysis of particular striations in the
corona, we characterized the evolution of image structure across the
HI1 field of view using structure functions (\citealt{SchulzRehberg1981,CoverThomas1991}),
which summarize the variability of a data set at different scales
and directions. Given a three-dimensional dataset such as our HI1
image sequence, $h(x,y,t)$, the second-order structure function is
a six-dimensional functional in both the independent variables and
their offsets:
\begin{equation}
S_{xyt}(x,y,t,\Delta x,\Delta y,\Delta t)\equiv\left(h(x,y,t)-h(x-\Delta x,y-\Delta y,t-\Delta t)\right)^{2}.\label{eq:structure-3d}
\end{equation}
$S_{xyt}$ is useful because it can be averaged across neighborhoods
of the primary variables' domains, to yield a location-dependent measure
of structure vs. scale in the datasets. We set $\Delta t$=0 and average
over $t$, to arrive at a four-dimensional time-averaged structure
function $\left<S_{xy}\right>$, which characterizes the average local
variability of the dataset as a function of location and scale:
\begin{equation}
\left<S_{xy}\right>(x,y,\Delta x,\Delta y)\equiv\left<\left(h(x,y,t)-h(x-\Delta x,y-\Delta y,t)\right)^{2}\right>,\label{eq:structure-2d}
\end{equation}
where the angle brackets represent averaging over time.  We carried
out the computation of the $\Delta t$=0 cut of $S_{xyt}$ at each
time and averaged the result across all HI1 frames, prepared as in
the middle panel of Figure \ref{fig:conformal-radial}, from 2008
Dec 15 to 2008 Dec 29, to yield $\left<S_{xy}\right>$
throughout the dataset with image $x$ ranging over azimuthal angle
$\alpha$, and image $y$ ranging over log($\varepsilon)$. To further
reduce noise, we smoothed the computed values of $\left<S_{xy}\right>$
by convolution in the $x$ and $y$ dimensions with a kernel {2.7\arcdeg}
wide in $\alpha$ and having the equivalent size in $\log(\varepsilon)$.
Further, to analyze radial evolution of ``typical'' striae and other
wind features without contamination from the potentially anomalous
central stripe that is visible at $\alpha=270^{\circ}$,
we selected two azimuths, $\alpha=250^{\circ}$
and $\alpha=290^{\circ}$ on opposite sides of the {270\arcdeg}
line, and averaged $\left<S_{xy}\right>$ between them. This further
reduced the dataset to 3 dimensions:
\begin{equation}
S(\varepsilon,\Delta\alpha,\Delta(\log(\varepsilon)) \equiv
\overline{
\left< S_{xy}\right> ( \{ 250^{\circ}, 290^{\circ} \} ,
\varepsilon , \Delta\alpha ,
\Delta ( \log( \varepsilon ))),  }
\label{eq:structure-1d}
\end{equation}
where the overbar represents the averaging over neighborhoods in $\alpha$
and $\varepsilon$. We used the two separate values of $\alpha$ to
symmetrize $S$ as a function of $\Delta\alpha$, while still eliminating
the structure near {270\arcdeg}.

\begin{figure*}
\epsscale{1.05}
\plotone{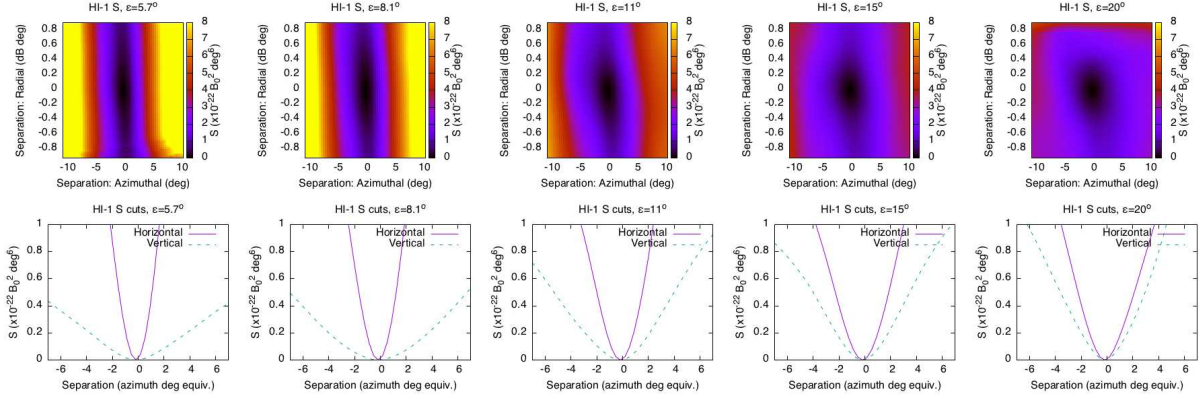}
\caption{\label{fig:2-D-structure-functions}2-D mage structure functions and
their cuts, averaged the entire processed HI1 dataset from 2008 Dec
15 to 2008 Dec 29, reveal the isotropization of image texture with
increasing altitude. See text for discussion.}
\end{figure*}

$S$ thus contains a separate image in the independent variables $\Delta\alpha$
and $\Delta\varepsilon$ for each separate value of $\varepsilon.$
Several of these structure images are shown in Figure \ref{fig:2-D-structure-functions}.
The transformed data use a logarithmic scaling for $\varepsilon,$
so these corresponding structure images are most naturally measured
in degrees of azimuth and decibels (dB) of elongation angle. One dB
corresponds to a factor of roughly 1.26 in $\varepsilon$, so that
in the $\varepsilon=11^{\circ}$ panel at center of Figure
\ref{fig:2-D-structure-functions}, a value of 0.8 dB in the radial
separation corresponds to a separation of {2.2\arcdeg}. Because
the underlying images are conformal to the original sky plane, the
structure function planes are also conformal and any apparent anisotropy
reflects a real anisotropy in the original images. 

Each image plot across the top row of Figure \ref{fig:2-D-structure-functions}
shows a single 2-D slice of \emph{S}: it is the squared difference
(averaged across azimuthal selections and time) between the value
of a single pixel at a given apparent radius and each selected azimuth,
and the corresponding value of nearby pixels with the corresponding
offset throughout the plotted image. The image values drop to zero
at the origin, because the average squared difference between a single
pixel and itself is identically zero. The tall, narrow structure of
the trough at central radii of {5.7\arcdeg}, {8.1\arcdeg}, and
{11\arcdeg} reflects the striation of the ``typical'' corona
at those altitudes: there is more variation of image value in the
$\alpha$ direction than in the $\varepsilon$ direction. The more
rounded structure at central radii of {15\arcdeg} and {20\arcdeg}
reflects the increasing isotropization of the brightness variation,
at higher altitudes. 

The plots in the bottom row of Figure \ref{fig:2-D-structure-functions}
are horizontal and vertical cuts through the origin of the 2-D structure
function slices. They reveal the increasing isotropy of the image
as the wings of the vertical cuts rise with altitude and the wings
of the horizontal cuts drop with altitude. It is immediately apparent
by inspection that the shift in the structure function is not due
to residual noise in the images: uncorrelated noise would produce
a sharp rise from the origin, rapidly transitioning to a shallower
slope at scales dominated by image features rather than by background
noise. The cuts grow monotonically with distance from the origin,
and show no sign of a noise-related break at the scales of interest.
The horizontal cuts grow more shallow with altitude, indicating fading
of the radial striae, even as the radial cuts grow steeper.

Note that we have followed the typical definition of $S$, which is
asymmetric rather than being symmetrized around a central point. This
means that $S$ can in principle be, and in practice is, asymmetric
with respect to the $\Delta$ variables at particular locations. For
example, there is a slight systematic asymmetry visible in the vertical
(radial) structure functions in Figure \ref{fig:2-D-structure-functions}.
This is because there is a systematic variation in the structure function
with $\varepsilon$, and point pairs taken in the positive-going $\Delta\varepsilon$
direction are centered at a different location than point pairs taken
in the negative-going $\Delta\varepsilon$ direction.

\begin{figure*}
\epsscale{1.05}
\plotone{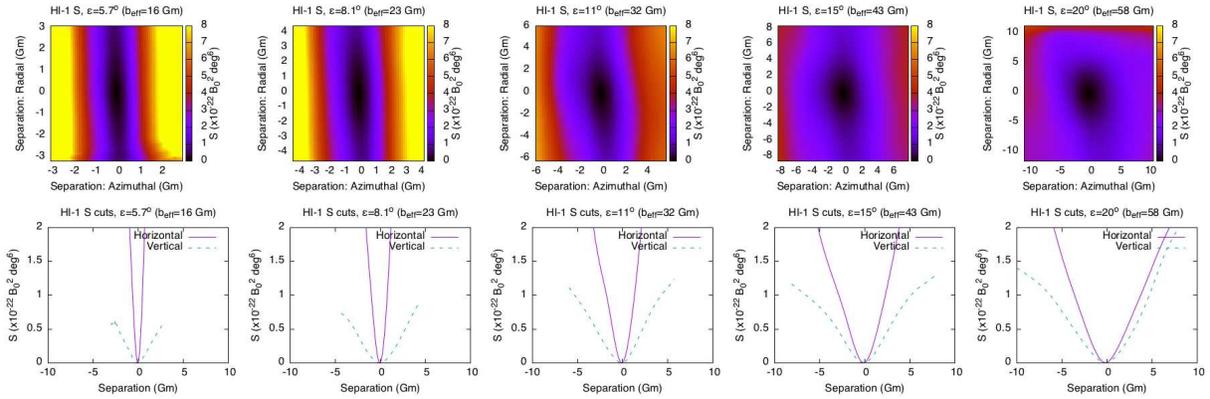}
\caption{\label{fig:2-D-quasi-spatial-structure}2-D quasi-spatial structure
functions and their cuts, averaged across the entire processed HI1
dataset from 2008 Dec 15 to 2008 Dec 29, reveal shifts in the solar
wind variability with radius from the Sun.}
\end{figure*}

To better understand the radial evolution, we also plotted the structure
function in quasi-spatial coordinates. These are plotted in Figure
\ref{fig:2-D-quasi-spatial-structure}. We made use of the known but
nontrivial observing geometry to make the conversion, which is somewhat
approximate. In particular, there is a variable ($\pm$5\% to $\pm$40\%)
cross-scale mixing effect, increasing with $b_{eff}$ across the field
of view, due to the shifting perspective along varying lines of sight.
Further, because we are observing a large fraction of the corona rather
than a compact feature such as a CME, the conversion from elongation
$\varepsilon$ to spatial distance from the Sun uses an ``effective
impact parameter'' $b_{eff}$, rather than the simple trigonometric
impact parameter $b$, for each line of sight. Both of these effects
are described at length in the Appendix.

\begin{figure}
\epsscale{1.15}
\plotone{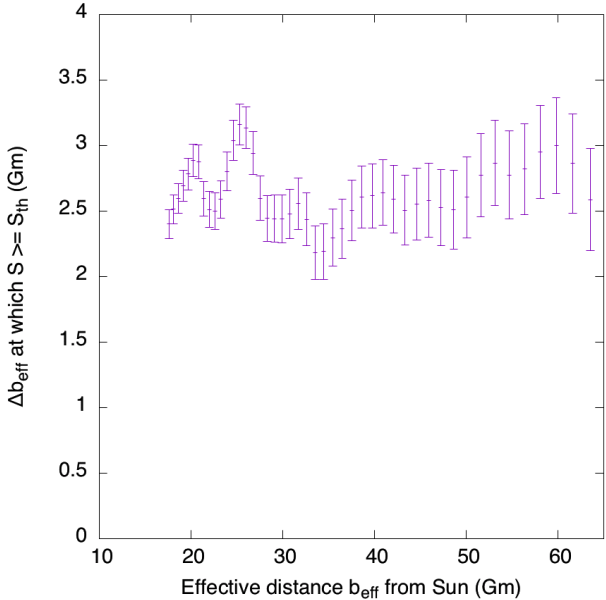}
\caption{\label{fig:distance-evolution}Plot of the separation distance $\Delta b_{eff}$,
for the radial cut of the structure functions in
Figure \ref{fig:2-D-quasi-spatial-structure}
to exceed a threshold $S_{th} = 0.5 \times 10^{-22}B_{\odot}^{2} \mbox{deg}^6$,
measures softening of the structure function with altitude. The radial
structure function of the images grows neither harder nor softer across
the effective altitude range of this study. Error bars are based on
a one-image-pixel offset in $\Delta b_{eff}$.}
\end{figure}

While the horizontal (azimuthal) structure function in the images
clearly expands faster than the overall radial expansion of the solar
wind (Figure \ref{fig:2-D-structure-functions}), the vertical (radial)
structure function in the images appears roughly constant as a function
of $b_{eff}$. Figure \ref{fig:distance-evolution} shows the evolving
trend. Because, as we demonstrated in Section \ref{sub:Feature-propagation},
the average scaled brightness of the wind does not change with altitude,
the separation distance $\Delta b_{eff,th}$ for $S$ to exceed a set
threshold $S_{th}$ is a valid measure of the changing hardness of
$S$ with solar distance. That is plotted in Figure \ref{fig:distance-evolution},
for $S_{th}=0.5\times10^{-22} \, B_{\odot}^{2}\mbox{deg}^6$. Throughout the
measurable range of offsets from the Sun, this thresholding distance
remains approximately constant: the radial cuts do not soften as do
the lateral cuts. This behavior is consistent with fully developed
turbulence in the radial direction.

\section{Discussion}
\label{sec:Discussion}

We have identified and characterized a fundamental change in the apparent
texture of the outflowing plasma above the top of the solar corona.
Between about {10\arcdeg} ($\sim$40 $R_{\odot}$) and about {20\arcdeg}
($\sim$80 $R_{\odot}$), the familiar radial striae of the solar
corona, which comprise the streamers, pseudostreamers, plumes, and
other radial structure in the corona, gradually fade until their brightness
is dominated in the slow solar wind {20\arcdeg} from the Sun by
more compact, nearly isotropic features (``flocculation''). We have
traced individual features in the flocculated visual field and found
that, unlike the Sheeley blobs and Viall puffs previously studied
by several other groups, which clearly originate in the corona (e.g.,
\citealt{SheeleyEtal1997,SheeleyRouillard2010,ViallEtal2010,ViallVourlidas2015}),
these flocculae largely ``fade in'' over the course of several apparent
degrees of propagation above {10\arcdeg} from the Sun. Both types
of visual feature co-exist in the outer field and could represent
either different physical phenomena or different parts of a single
distribution of fluctuations in the solar wind.

\subsection{Fading of the Striae}
\label{sub:Fading-of-the}

The striae themselves fade faster with respect to radius than would
be expected from bright features in a constant-speed, conservative
solar wind, eventually dropping below our processed images' noise
floor---where they can be seen by further, more aggressive spatial
averaging of the HI1 images to further beat down the noise floor,
at the cost of further reducing spatial resolution. There is no large-scale
breakup of the striae as might be expected from large-scale instabilities.

We are confident in this fundamental observational result, because
we have eliminated the most plausible observational scenarios that
might mimic fading of the coronal striae. In particular, by demonstrating
that the striae, while faint, can be observed even under the reduced
noise floor from processed HI1 data, we have shown that noise and
background effects alone are not responsible for the apparent fading
of the striae---and this result is corroborated by the opposite
brightness trend in the flocculae, which have similar spatial scales and
scaled brightness values to the striae themselves. 

\new{The striae are rendered visible by Thomson scattering, and therefore
the features' geometry relative to the Thomson surface---the locus
of the sphere whose diameter extends from the observer
to the Sun---becomes important.
The Thomson surface marks the point of greatest 
apparent surface brightness per unit density along a given line of sight, 
and therefore purely radial structures may gain or lose apparent surface brightness
relative to the applied $\varepsilon^{-3}$ scaling, depending on this 
geometry.  In fact \citet{VourlidasHoward2006} proposed to use this effect
to locate features in 3-space by looking for anomalous dimming as the features
propagate.  However,
the Thomson surface is surrounded by the ``Thomson plateau''
(\citealt{HowardDeForest2012}): a broad range of out-of-image-plane angles over
which feature radiance is very nearly independent of geometry, on a given line of sight.}

We eliminate the possibility that the fading is due to scattering physics, 
by noting that no striae appear to fade in with altitude, only to
fade out. Because our dataset spans over half a solar rotation and
striae are observed to exist across a wide range of coronal longitudes,
Thomson plateau effects should cause cross-fading between different
striae and not a uniform attenuation of all striae. That is because,
in the overall population of dense radial features, approximately
the same number of radial features should be outside the Thomson plateau
at low altitudes and enter it at high altitudes, as exist
within the Thomson plateau at low altitudes and exit it at high altitudes.
\new{Moreover, because the field of view is smaller than the angular extent
of the Thomson plateau, it is expected that most
visible striae remain within the plateau for their entire visible length.}

Turning to more subtle solar geometric effects: the ``pancake effect''
of azimuthal expansion in the solar wind, coupled with perspective/projection
effects, can cause fading of features that are compact in the lower
corona and separated radially. Features of that description can apparently
merge in the image plane as they propagate outward, even when no physical
merging or mixing occurs; this is illustrated with a simple geometric
argument by \citet{ViallEtal2010}. Similar effects can cause apparent
merging in blobs that pass one another along the line of sight, as
quantified for the `Sheeley blobs' by \citet{SheeleyRouillard2010}.
These effects impact variations along the radial direction and may
affect the ``flocculated'' appearance. Importantly for the current
work, these effects work to blend individual features and produce
an illusion of a radial stria, rather than the opposite, so they cannot
account for the general fading of the striae. To cause a stria to
fade, this mechanism requires a very rare coincident co-alignment to
cause an illusive stria close to the Sun and reveal puffs farther
from the Sun.

The fading of the striae, in the scaled brightness parameter
$B_{\odot} \, \mbox{deg}^{3}$,
is surprising because it varies from expected behavior of a mass-conserving,
constant-speed radial solar wind. The two most obvious physical mechanisms
to explain it are (A) that the wind might not be constant speed, or
(B) that the features themselves might not conserve visible mass. 

We eliminate acceleration effects (possibility A) by noting that,
to explain the observed fade of a factor of more than 3 in scaled
brightness, the wind inside the striae would have to accelerate by
a comparable factor across the field of view. No such acceleration
is measured, and the outflow speed of bright features varies by less
than a factor of 2 across the entire image (and considerably less
along a given radius). 

Turning to conservation effects (possibility B): conservation of bright
structure is not necessarily conservation of mass. The images are
produced using sunlight that is Thomson-scattered off of free electrons.
Photometric mass estimates in the heliosphere (\citealt{DeForestEtal2014})
or corona (\citealt{VourlidasEtal2000}) are produced by assuming
the free electrons are part of a neutral, 100\% ionized plasma with
approximately coronal composition (specifically, proton-to-helium
ratio). Recombination of ions and free electrons could in principle
cause the striae to fade, but this effect would require more than
2/3 of the plasma to neutralize. That effect is implausible: although
electrons in the solar wind typically suffer a few tens of collisions
enroute from the Sun (\citealt{Salem2003}), by the altitude range
under consideration (40--100 $R_{\odot}$), collisions are very rare
(\citealt{HundhausenEtal1968,OwockiScudder1983}). Further, the high
kinetic temperature of order $10^{5}$--$10^{6}$~K prevents recombination
(\citealt{Gibson1973}). 

Eliminating (A) and recombination effects, leaves the possibility
of intrinsic variation of the entrained mass within each stria. Such
mass loss would be due to diffusion or other effects that transport
material from the dense structure to the intervening spaces.

\new{One possibility for structural mass loss is misalignment, at
high altitudes, of quasi-planar
structures that may, by coincidence, align with the line of sight when seen closer to the Sun.  This 
possibility is important because many streamers and pseudostreamers are thought
to have quasi-planar geometry rather than being compact structures
(\citealt{WangEtal2007}).
But the same argument applies to these effects as to effects from the geometry of 
Thomson scattering: if the striae were primarily due to chance alignments with 
quasi-planar density structures (and loss of alignment were the reason most observed striae fade at
large radii), then chance alignments at those farther distances 
should be expected to cause a roughly
equal number of striae to fade in as out --- which is not observed.  We conclude that the fading striae are not  
due to chance alignments and subsequent misalignments of smoothly expanding quasiplanar density
structures in the solar wind.}

Our discussion
so far has focused on eliminating various observational effects that
could in principle explain the apparent radial evolution seen in the
images and diagnostic parameters shown in Section \ref{sec:Results}.
It remains to offer a physical scenario to explain the fading of striations
and emergence of puffs and flocculation, and the isotropization of
the computed structure functions of the images. We now consider the
effects of the turbulent fluctuations observed at 1~AU, extrapolated
inward to the observed range of altitudes, and whether they can account
for the inferred lateral transport of mass.

We begin by recalling the general changes in nonlinear behavior moving
outward from the Sun. Below the Alfv\'{e}n surface where the solar wind
speed exceeds the radial speed of fast-mode MHD waves, the plasma
is also low $\beta$ and dynamics are strongly influenced by the magnetic
field. This gives rise to a strong correlation (or spectral) anisotropy
relative to the magnetic field direction (\citealt{ZankMatthaeus92,CranmerEtal2015,OughtonEtal2015}).
Under these conditions, magnetic flux tubes strongly align with the
large scale magnetic field, and magnetic \old{fluxtuation}\new{fluctuation} gradients concentrate
in the perpendicular direction. The variability of density across flux
tubes then gives rise to strong anisotropic density gradients that,
in turn, appear as striae in the imaging data. In the sub-Alfv\'{e}nic
coronal regions, the dynamical emergence of this anisotropy is in
part due to active anisotropic spectral transfer (\citealt{ShebalinMatthaeusMontgomery1983,OughtonEtal1994})
that amplifies the transverse gradients relative to the radial ones.
Furthermore, in the corona this turbulence is possible because inward
propagating fluctuations overcome the outward flow of the solar wind
and interact with their outward propagating counterparts (MHD wave-wave
interactions).

Above the Alfv\'{e}n surface, both of these effects change. First, inbound
(in the co-moving frame) waves are advected outward (in the solar system
stationary frame), reducing the overall strength of the wave field
and delaying development of active Alfv\'{e}nic turbulence. Second, above
the related $\beta=1$ surface the magnetic field strength is no longer
dominant over either thermal or convective effects. On the other hand,
the state of the plasma flowing into this critical region near the
Alfv\'{e}n and $\beta=1$ surfaces is already highly anisotropic, having
been shaped by the well-ordered and nearly radial average magnetic
field. This accounts for the dominant appearance of striae, interrupted
by discrete features including CMEs, etc., in images from coronagraphs
and our HI1 dataset. Therefore, the observed striae seen at or above
the $\beta=1$ surface are consequences of factors operating at lower
altitudes. Although the striae persist into our field of view, the
factors that build and maintain the coronal anisotropy are no longer
present in the young solar wind.

One may now ask whether the existing turbulence field in the solar
wind at 150 Gm (1~AU), extrapolated inward to 38~Gm ($b_{eff}$
for lines of sight close to {12\arcdeg} from the Sun), account
for the breakup of the striae that we observe there. At 150 Gm, typical
values of the turbulent velocity $Z(150$~Gm$)$ and correlation
length $L$(150~Gm) are 25~km~s$^{-1}$ and $1$~Gm, respectively
(\citealt{RuizEtal2014}; \citealt{IsaacsEtal2015}). Both of these
quantities are approximately log-normally distributed, so that extreme
outliers are to be expected. Observations suggest that $Z\propto r^{-1/2}$
in the inner heliosphere (\citealt{VermaRoberts1993}), so that
$Z$(38~Gm)~$\approx 12$~km~s$^{-1}$.
Further, taking the correlation length to scale between $L\propto r$
and $L\propto r^{1/2}$ (\citealt{BreechEtal2008}), at 38~Gm (0.25~AU)
the correlation scale should be $L(0.25AU)\approx0.25$~Gm \textendash{}
0.5~Gm if the turbulent field observed at 150 Gm (1~AU) persists
inward to this extent. 

Based on the estimates above, correlated density or flow features
in the outer reaches of the solar wind observed in Figure \ref{fig:conformal-radial}
have an estimated expected size of 0.25--0.5 Gm, and subtend approximately
{0.25\arcdeg}--{0.5\arcdeg} of azimuth. Features of this scale
are close to the resolution limit of the present observation, but
large-scale outliers, extending to several times the expected correlation
scale, should be observed if the the near-Earth turbulent field happens
to be already well developed in the observed range of altitudes.

\begin{figure}
\epsscale{1.11}
\plotone{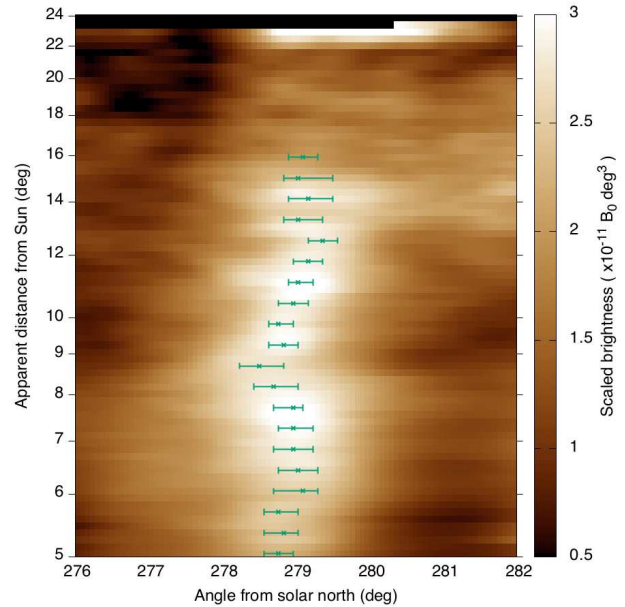}
\caption{\label{fig:Comoving-evolution-plot}Co-moving evolution plot shows
brightness in a horizontal cut through the stria in the bottom row
of Figures \ref{fig:Evolution-sequences-within}
and \ref{fig:Smoothing-the-final},
in a co-moving frame rather than at a single time. The overlain curve
and error bars indicate the location of the brightest portion of the
stria. The apparent shifts in azimuth are marginally inconsistent
with pure radial motion, suggesting lateral perturbations to the outward
motion. See text for details.}
\end{figure}

In fact, such displacements appear to be observed. Figure \ref{fig:Comoving-evolution-plot}
shows the lateral evolution of the stria in the bottom row of Figures
\ref{fig:Evolution-sequences-within} and \ref{fig:Smoothing-the-final}
as it propagates outward; this stria is \old{appears} typical and several
other striae show similar behavior. The image shows the average brightness
evolution, averaged vertically across 10\% of image height, in a horizontal
band extending across each of the subfield images in the Figure \ref{fig:Evolution-sequences-within}
dataset. The overlain points reveal lateral evolution of the center
of the stria in the co-moving frame of the solar wind. Each point shows
the horizontal location of maximum brightness at the corresponding
time and apparent distance (elongation $\varepsilon)$. We determined
the error bars by noting that the RMS noise level
at $\varepsilon=14^{\circ}$
is $4 \times 10^{-13} \, R_{\odot}$~deg$^{3}$, and finding the location
where the radiance in each horizontal cut falls from the maximum by
at least that amount. The points are calculated only for every third
frame of the dataset, to avoid the effects of the three-frame temporal
averaging we used during the processing: each point represents a true
independent sample.

While the variations in azimuth do not exceed the error bars for most
pairs of points, the coherence of the lateral motion renders the motion
statistically significant. The amplitude is approximately {0.5\arcdeg}
in azimuth, in good agreement with the estimate above.

The lateral motions that we observe are inconsistent with direct radial
motion and consistent with either scaling of the fully developed turbulence
detectible near Earth, with local onset of lateral turbulent motion
due to hydrodynamic instabilities (such as the Kelvin-Helmholtz instability),
or with isotropization of MHD turbulence present in the corona itself.
Lateral convective transport from these processes must occur to account
for the fading of the striae, but there is no clear and direct sign
of the postulated breakup. Therefore, we conclude that the breakup
must occur on scales smaller than the effective spatial resolution
of $\sim 0.5$--2.5~Gm.

\subsection{Onset of the Flocculation}

In counterpoint to the fading of the striae, flocculae ``fade in''
smoothly from elongation angles of $\sim${8\arcdeg} out to {20\arcdeg}.
The radial length scale of the flocculae is comparable to the length
scale in the Viall puffs and/or Sheeley blobs that enter the HI1 field
from the lower corona: 3--10~Gm. Two explanations spring immediately
to mind: (1) the flocculae could be local interaction regions: zones
of density enhancement caused by variation of outflow speed along
a given wind flow streakline\footnote{\new{Readers are reminded that a streakline is the path in space
traced by a particular test particle in a time-dependent flow field; this is distinguished from a streamline, which 
is the path in space traced by an incompressible flow field at a particular moment in time.}}; (2) the flocculae could be a symptom
of the same instabilities that are seen to offset and fade the striae.

To investigate the hypothesis that the flocculae are caused by local
interactions between different-speed streams, we consider the divergence
of the velocity field $\nabla\cdot\mathbf{v}$ necessary to cause
the observed growth rate of the density in particular flocculae (as
in Figure \ref{fig:Faint-relatively-compact}). To determine this
divergence, we must estimate the proportional density enhancement
represented by the brightness growth of the flocculae.

Because of the way that coronal and heliospheric images are created,
there is no ``absolute zero'' to unpolarized Thomson scattering
images in general. This is because the images are produced by subtracting
a steady or nearly-steady background model in the fixed focal plane
(coronagraphs) and/or the Celestial sphere (heliospheric images);
and there is no intrinsic difference between a steady signal from
the F corona or stray light, and a steady signal from a stationary
or quasi-stationary electron density structure in the field of view.
Hence, essentially \emph{all} unpolarized Thomson scattered images
from current instruments report ``feature excess'' brightness, and
hence inferred densities are only ``feature excess'' densities,
compared to an unknown steady background. 

Near Earth, the typical electron density of the slow solar wind is
3--10 cm$^{-3}$. Extrapolating inward to {10\arcdeg} elongation
from the Sun (a line of sight angle whose weighted average distance
from the Sun is 28~Gm; see the Appendix), we estimate that the ``typical
background'' electron density at {10\arcdeg} elongation in our
images is 100--300~cm$^{-3}$, along an effective line-of-sight distance
of 50 Gm (0.33~AU). Taking the lower figure and following Equation
5 of \citet{HowardDeForest2012}, we infer that 
\begin{equation}
B_{bk} (\varepsilon=10^{\circ}) \approx
\frac{B_{\odot}\sigma_{t}\pi r_{\odot}^2}
{((1AU)(\sin(10^{\circ}))^2} \, (100 \, \mbox{cm}^{-3})
(50 \, \mbox{Gm}) \, ,
\label{eq:brightness-from-density}
\end{equation}
where $B_{bk}$ is the calculated background radiance from Thomson
scattering, $B_{\odot}$ is the mean solar radiance, $\sigma_{t}$
is the Thomson scattering cross section of $4.0\times10^{-30}$~m$^2$,
and $r_{\odot}$ is the mean solar radius of $6.96\times10^8$~m.
This evaluates to
$B_{bk}(\varepsilon=10^{\circ})=4.2\times10^{-14}B_{\odot}.$
Multiplying by $\varepsilon^{3}$ yields the elongation-independent
scaled background radiance $B'_{bk}=4.2\times10^{-11}B_{\odot} \mbox{deg}^3$.
This $B'_{bk}$ remains approximately equal throughout all values
of $\varepsilon$ that we observed.

$B'_{bk}$ is comparable to the measured feature-excess brightness
in typical flocculae as seen in the rightmost panels of Figure \ref{fig:Faint-relatively-compact}.
The calculated $B'_{bk}$ exceeds this feature brightness by a factor
of 3--10. However, a typical floccule does not extend along the entire
line of sight. Taking the flocculae in Figure \ref{fig:Faint-relatively-compact}
to have roughly the same extent out of the screen as in azimuth, they
subtend roughly {10\arcdeg} relative to Sun center, rather than
the approximately {78\arcdeg} subtended by the effective line
of sight---and therefore the excess density associated with each
floccule is larger than the calculated ratio of
$B'_{floccule}(\varepsilon)/B'_{bk}$
by a factor of $\sim$8. Hence the floccule's excess density is roughly
equal to the density of the background flow at the same positions
on the line of sight, at the highest elongations in
Figure \ref{fig:Faint-relatively-compact}
($\sim${20\arcdeg} from the Sun). This implies that,
between altitudes of roughly 40 $R_{\odot}$ and 80 $R_{\odot},$
the density in a floccule approximately doubles compared to quiescent
propagating wind.

The flocculae have typical radial sizes of 3--6 Gm, so if the density
enhancement arises from a simple negative divergence of the bulk flow,
the material at the visible edges must propagate inward relative to
the mean bulk flow, crossing the width of the floccule during a transit
across 40 $R_{\odot}$ (30 Gm). Thus, the average $\Delta v$ between
the leading and trailing sides of the floccule must be about 10\%-20\%
of the wind speed, corresponding to a $\pm$5\% to $\pm$10\% antisymmetric
deviation from the mean radial outflow speed in the vicinity of each
floccule.

In light of the existence of highly structured radial striae, cross-field
evolution due to instabilities or onset of turbulence must play a
large role in the development of the flocculae. If the flocculae were
generated by purely compressive processes due to temporal variation
in the speed of individual flux tubes' coronal wind, as described
above, then their azimuthal size should be comparable to the observed
scale of individual wind structures---i.e., striae. Instead, we observe
the flocculae to be, typically, over  3 times larger in azimuthal
extent than are typical striae. This implies collective behavior on
scales larger than the natural azimuthal scale on which the wind is
injected from the corona itself. 

An alternative explanation is that the flocculae are density enhancements
caused by turbulent perturbations to the flow field rather than by
pileup in a smooth flow. In this view, turbulent motions within the
flow give rise to density perturbations on length scales comparable
to the correlation scale of the turbulence. We note, however, that
the radial length scale of the flocculae is about 10$\times$ larger,
and their azimuthal length scale is perhaps $30\times$ larger, than
the scaled correlation length calculated in Section \ref{sub:Fading-of-the}.
In this regard, one might explain the flocculae as density enhancements
associated with the very largest and strongest velocity perturbations,
associated with the onset of turbulence in the inner heliosphere.
The moderately frequent occurrence of dynamically active fluctuations
appears to be reasonable, based on the observed long tails on the
distributions of correlation lengths (approximately log normal) seen
in {\em Helios} data (\citealt{RuizEtal2014}). It is also possible that
the flocculae emerge as a manifestation of so called ``$f^{-1}$''
noise, which is observed in situ at very low frequencies in solar
wind density observations (\citealt{MatthaeusEtal2007}).

A detailed analysis of how and whether exceptional turbulent fluctuations,
$f^{-1}$ noise, or some other collective phenomenon may account for
the present observations is a topic for future work.

\section{Conclusions}
\label{sec:Conclusions}

We have imaged, above the solar corona, a fundamental shift in texture
of the solar wind: from the highly anisotropic, magnetically structured
coronal plasma to more isotropic, flocculated solar wind plasma. We
have eliminated several of the more common sources of systematic error,
including image degradation due to the weaker signal at increasing
solar elongation, and conclude that the transition is real. We observed
the ultimate disposition of the radial striae that, aside from the
rapid falloff as the cube of elongation angle $\varepsilon^{3}$,
are the most prominent visual feature of the outer corona. Further,
we have noted and described the simultaneous development of flocculae:
local density enhancements that arise in the outflowing solar wind
as the striae disappear. These two effects together combine to radically
change the visual texture between the outer corona and the inner heliosphere,
and mark the profound shift between the primarily magnetically structured
corona and the primarily hydrodynamic solar wind. 

Moreover, the shift in texture points strongly to the early development
of a turbulent cascade, of which we can observe primarily the energy-bearing
and largest inertial scales. The behavior of the features is consistent
with onset and development of quasi-isotropic turbulence due to hydrodynamic
and MHD instabilities, and inconsistent with smooth flow.

The striae fade gradually at resolutions attainable with the HI1 measurement
at apparent distances (elongation angles) of {10\arcdeg}--{20\arcdeg}
from the Sun, corresponding to 44--88 $R_{\odot}$ in effective line-of-sight
impact parameter $b_{eff}$. In that interval they typically fall
in brightness by a factor of $\sim3$--5 compared to the behavior
expected for a smooth, unperturbed solar wind. In that same interval,
they undergo slight non-radial distortions that, while only marginally
detected, are \emph{prima facie }consistent with the large-scale tail
of a forming turbulent cascade whose parameters scale correctly to
the observed variability of the slow solar wind 150 Gm (1~AU) from
the Sun. The striae, while faint, still exist at the outer edge of
the HI1 field of view (solar elongation of {24\arcdeg}, off the
centerline of the images), and become visible with broader smoothing
of the images. 

In approximately \new{the} same altitude range that the striae fade, localized
high-density puffs of solar wind (which we term ``flocculae'') develop
gradually, fading into existence without noticeable lateral spreading
or internal motion as they brighten. The flocculae are consistent
with variations in local wind speed of the order of $\pm$5\%--15\%,
but require collective motion to form at the azimuthal size scales
observed. This, too, may be a signature of either developing turbulence
or (equivalently) the early onset and action of large-scale hydrodynamic
instability (e.g., \citealt{Ofman2016}).

The disappearance of the striae, and appearance of the flocculae,
are not due to Thomson surface projection effects.  The entire population of striae fades out, and
the entire population of flocculae fades in; 
Thomson effects would cause some of each type of feature to fade out and others
to fade in with altitude, without (on average) changing the total populations. 
Contrariwise, the \new{nuances of Thomson scattering at
high solar altitudes yield a} calculated ``Thomson plateau''
of nearly constant apparent radiance per unit density
(\citealt{HowardDeForest2012}).
This plateau is quite broad---over {100\arcdeg} in
out-of-plane angle relative to the Sun---compared to
the field of view of our measurements.  Therefore, most
striae and most flocculae remain
well within the plateau throughout the observed range of apparent
distances. 

In producing the images, we have demonstrated several novel-to-coronagraphy
image processing techniques: minsmooth unsharp masking, time-shifted
image co-addition, and structure function analysis to characterize
ensemble texture. These techniques are almost certainly useful for
related studies well beyond this simple analysis.

This study is close to the limit of what is possible with the existing
HI1 instruments. Higher sensitivity, higher resolution (in space and
time) images of this transition zone from coronal to heliospheric
physics are necessary to reveal whether, in fact, formation and development
of a nearly isotropic turbulent cascade is responsible for the loss
of coronal structure and development of variable flocculated structure
in the solar wind at these large observable scales, to and identify the
relative importance of coronal features and instability growth to
the structure of the solar wind itself. The upcoming Solar Probe mission
will reveal wind structure on fine scales and can measure parameters
such as the Alfv\'{e}n speed, wind speed, and $\beta$ parameter directly
at particular locations---but capturing the effect of these transitions
and their import for the energy-bearing range of solar wind turbulence
will require both these insights and a more global perspective on
the structure of the corona to solar wind transition.

\acknowledgements

C.E.D.\  is supported by grant NNX15AB72G from NASA's Living With a
Star Targeted Research \& Technology program.
W.H.M.\  is supported by the 
Solar Probe Plus IS$\odot$IS project,
NSF SHINE grant AGS-1156094, the NASA LWS
grant NNX15AB88G, and the Heliospheric GCR grant NNX14AI63G.
N.M.V.\  is supported
by the NASA's Heliophysics Guest Investigator program.
S.R.C.\  is supported by NSF SHINE grant 
AGS-1540094, NASA HSR grants NNX15AW33G and NNX16AG87G, and start-up funds from the 
Department of Astrophysical and Planetary Sciences at the University of Colorado, Boulder.
The work was
improved by informative discussions with, and acerbic commentary from,
T.\  A.\  Howard. The authors gratefully acknowledge the \emph{STEREO}
team for making their data available to the public, and the Solar
Probe Science Working Group for making it clear that this article
was necessary. The analysis relied heavily on the freeware Perl Data
Language (http://pdl.perl.org).

\appendix

\section{A. Geometry and Spatial Scales}

\begin{figure*}
\epsscale{1.00}
\plotone{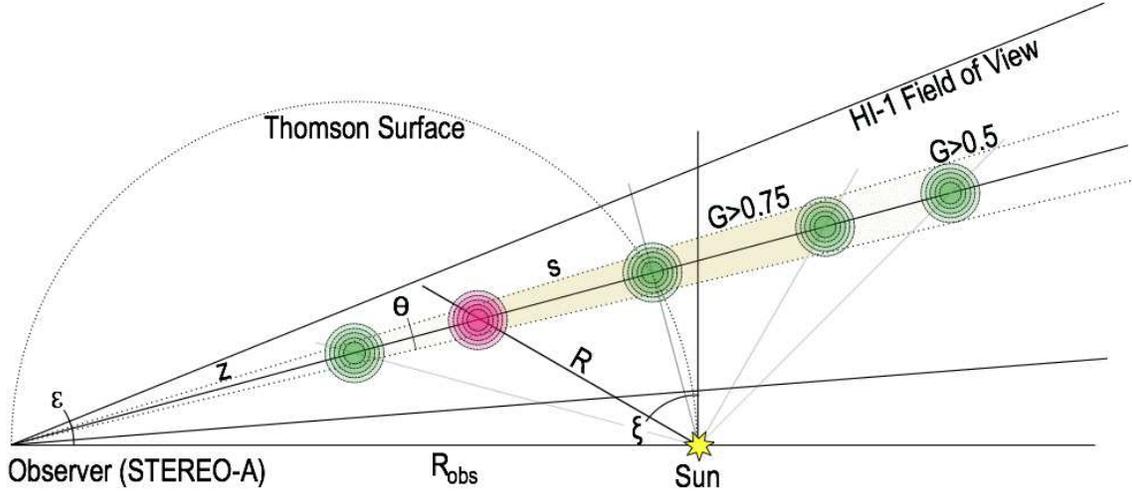}
\caption{\label{fig:Perspective-diagram-shows}Perspective diagram shows the
difficulty of converting from (angular) image scales to spatial scales
in the structure function. The geometrical factor $G$ relating density
to radiance is nearly flat for a range of approximately {90\arcdeg}
in the out-of-plane angle $\xi$ along a given line of sight. As elongation
angle $\varepsilon$ increases, the Thomson plateau grows sufficiently
broad that wind structures from many different spatial scales contribute
to image features of a given apparent size.}
\end{figure*}

Analyzing the spatial structure function of the corona, as sampled
by our images, requires converting the scaled brightness images from
their native angular coordinates to distance units in the corona.
This is non-trivial, because the scales sampled by the image and its
structure function in the vicinity of a given pixel are sampled in
observer-centered angular coordinates, relating variation of the brightness
along an integrated line of sight to that line of sight's azimuth
and elongation angles ($\alpha,\varepsilon)$ relative to the observer
and Sun. In contrast, the structure function of interest to understand
turbulent onset is sampled in spatial coordinates, relating variation
in density (or its proxy, Thomson-scattering emissivity) to spatial
offset, as a function of radial distance \emph{r} from the Sun. The
various angles and distances are shown in Figure \ref{fig:Perspective-diagram-shows};
and $\alpha$ is considered to be an angle of revolution about the
observer-Sun line.

For this initial treatment of structure function variation, we treated
the neighborhood of a structure function sample as a collection of
plane-parallel lines through a finite locus of corona, to approximate
the more complex geometry of the sampling. This introduces some ``slop''
in the scales represented by sample offsets ($\Delta x$,$\Delta y$)
in the spatial structure function, and in the radii from the Sun represented
by particular base sample location (\emph{x,y}) from Equation \ref{fig:2-D-structure-functions}.
Here we consider that treatment and explore the limits of the approximation
used.

The 2-D images from HI1 are integrals through the outer corona and
solar wind. To summarize the geometry, we introduce the angle $\xi$,
which is the angle between a given feature and the local ``sky plane''
at a given image location, relative to the Sun itself. This angle
is shown in Figure \ref{fig:Perspective-diagram-shows}. Over a very
wide range of $\xi$, called the ``Thomson plateau'' by
\citet{HowardDeForest2012},
radiance at the focal plane due to a particular density feature is
approximately independent of $\xi$. The Thomson plateau causes the
line-of-sight integral to extend over a range that is significant
compared to the observer-Sun distance $R_{obs}$ ($\sim 150$~Gm),
mixing both spatial scales and feature-sun distances in the image
plane. 

Both the scale and radial-offset effects depend on solar elongation
angle $\varepsilon$ of a given feature. At low $\varepsilon$, pixels
receive significant brightness contributions from a range of line-of-site
positions $\Delta s$ that is small compared to $R_{obs}$, while
at higher $\varepsilon$, each pixel receives brightness contributions
from a broader $\Delta s$ that is comparable to $R_{obs}$. So at
greater elongations from the Sun, more cross-scale mixing occurs. 

The other mixing effect of import is mixing across different values
of $r$. The reported brightness in each pixel is a brightness integral
along a particular line of sight. The integrand is significant along
a finite range of positions $s$ both ahead of and behind the
Thomson Surface. This means that most of the sampled wind along a
given line of sight is from locations $r(s)$ that are farther from
the Sun than $b$ (the impact parameter of the line of sight, i.e.,
the minimum value of $r$ along the line of sight). We consider
and quantify both these effects.

\begin{figure}
\epsscale{0.76}
\plotone{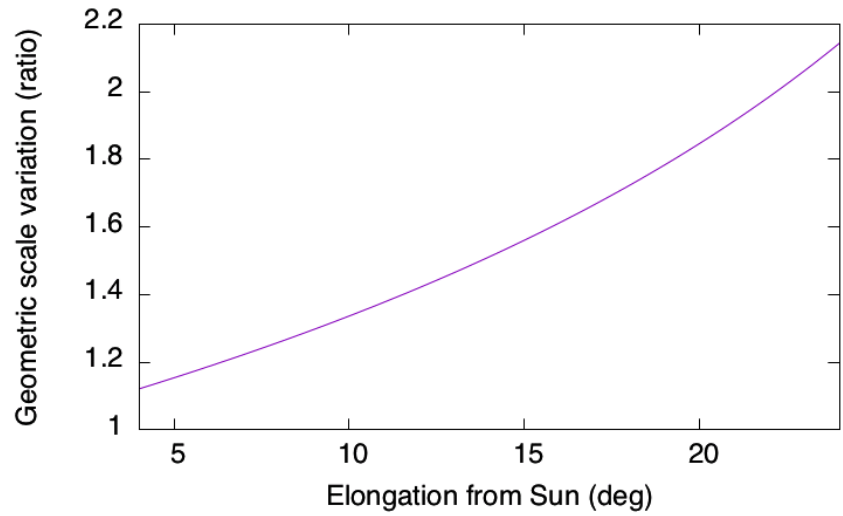}
\caption{\label{fig:Calculated-geometric-scale}Calculated geometric scale
ratio in the image 2-D structure function describes cross-scale smearing
against elongation angle $\varepsilon$. The image structure function
at {5.7\arcdeg} (left panels of Figure \ref{fig:2-D-structure-functions})
contains about $\pm$10\% variance in spatial scale, while the structure
function at {20\arcdeg}
(right panels of Figure \ref{fig:2-D-structure-functions})
contains about $\pm$40\% variance in spatial scale.}
\end{figure}

To quantify radial mixing and scale mixing, we identified the range
of sky angles $\xi$ that are important to a given pixel in the HI1
images. The Thomson plateau is defined by the geometric $G(\chi)$
effective per-electron scattering function described
by \citet{HowardDeForest2012}.
$G(\chi)$ mixes the illumination function and the inefficiency
of Thomson scattering at right angles, and is quite flat over the
range illustrated in Figure \ref{fig:Perspective-diagram-shows}.
But $G(\chi)$ is useful for comparing radiance of isolated
hypothetical features with similar density, and we are concerned instead
with a collection of wind features whose collective density falls
as $r^{-2}$. Because of this trend in density (Equation \ref{eq:density-scaling}),
the line of sight integral is better described as
\begin{equation}
B(\varepsilon,\alpha) =
k \int n_{e0}\frac{G(\chi(s,\varepsilon))}{(r(s,\varepsilon))^{2}}ds
\approx k \langle n_{e0} \rangle (b(\varepsilon))^{-2}
\int\frac{(b(\varepsilon))^{2}
G(\chi(s,\varepsilon))}{(r(s,\varepsilon))^{2}} ds \, ,
\label{eq:brightness-integral}
\end{equation}
where $B$ is the radiance observed along a given line of sight, $k$
is a scaling factor containing the various constants described by
\citet{HowardDeForest2012}, $n_{e0}$ is a radially-scaled electron
density, the angle brackets indicate line-of-sight averaging, the
split factor of $(b(\varepsilon))^{2}$ eliminates the $\varepsilon$
dependence of the integrand (because $r\propto b$), and the integrand
at right, which we call $G_{r}(\chi),$ is just the $G(\chi)$
described by \citet{HowardDeForest2012}, scaled by the known average
falloff rate of the solar wind density. $G_{r}(\chi)$ has no plateau
and it has a magnitude above 0.5 for only a {78.5\arcdeg} range
of angles, compared to the {114\arcdeg} range for $G(\chi)$.
Taking this range as the effective limit of the line-of-sight integral
for each pixel, the spread in spatial scales is approximately the
ratio of distances $z$ to the observer at the nearer and farther
limits of the angular range. The distance $z$ is just $R\sin(\pi-\varepsilon-\chi)\sin(\chi)^{-1}$
. The ratio of scales sampled in a particular small patch of image
therefore depends on the elongation $\varepsilon$ of that patch from
the Sun. The relationship is plotted in Figure \ref{fig:Calculated-geometric-scale}.
Over the range of elongations $\varepsilon$ considered in this work,
the ratio of the largest to smallest scale that affects the structure
function at a given apparent separation $\Delta\varepsilon$ or $\Delta\alpha$
reaches a maximum of about 1.8. This has the effect of smoothing out
high-polynomial-order variations in the image structure functions,
but does not---in first approximation order---affect the slope of
the nearly-linear sidewalls of the structure function cuts shown in
Figures \ref{fig:2-D-structure-functions} and
\ref{fig:2-D-quasi-spatial-structure}.

\begin{figure}
\epsscale{0.76}
\plotone{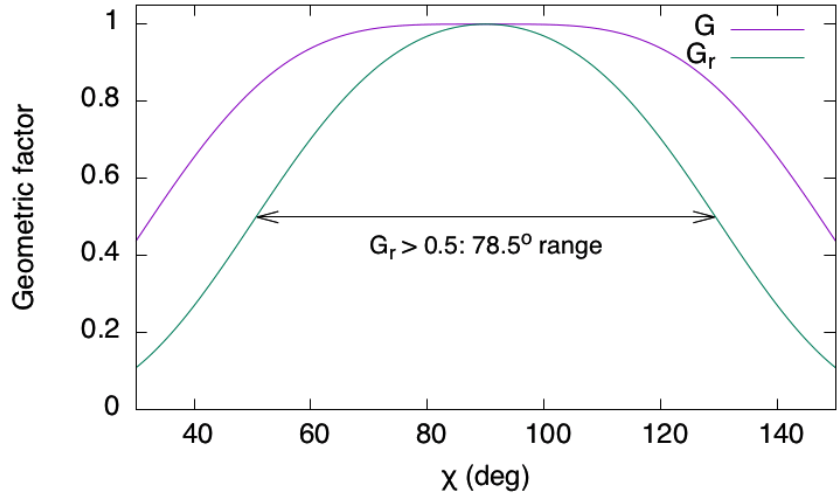}
\caption{\label{fig:The-modified-geometric}The modified geometric
term $G_{r}(\chi)$
is appropriate for determining the effective length of the line-of-sight
integral along each pixel of a HI1 image when viewing the distributed
background solar wind rather than a single bright isolated feature.
The {78.5\arcdeg} range of angles for which $G_{r}(\chi)>0.5$
is approximately two thirds the {114\arcdeg} range for which $G(\chi)>0.5$.}
\end{figure}

Similarly, integrating along the line of sight in Figure \ref{fig:Perspective-diagram-shows}
yields a weighted average of features sampled at different radii $r$
from the Sun. The most obvious effect is that the average solar distance
of a given line of sight is greater than the plane-of-sky distance
$b$. We name this average distance $b_{eff}$ and calculate it as:
\begin{equation}
b_{eff}=\frac{\int G_{r}(\chi)\sec(\tan^{-1}(s/b))ds}{\int G_{r}(\chi)ds}
\approx
\frac{\int_{\chi=50.75^{\circ}}^{\chi=129.25^{\circ}}
\sec(\tan^{-1}(s/b))ds}{1.63}\approx1.09\,b,\label{eq:beff}
\end{equation}
where the approximation treats the integral as a simple average over
the FWHM of $G_{r}(\chi)$. So the radial effect results in a line-of-sight
$b_{eff}$ that is just under 10\% higher than the calculated sky-plane
distance \emph{b} itself; and the range of radii sampled along a given
line of sight is approximately $\pm$10\% about the expanded $b_{eff}$.

This $b_{eff}$ is the line-of-sight impact parameter value that we
used in converting angular (image) values to spatial values, for Figure
\ref{fig:2-D-quasi-spatial-structure} and subsequent discussion.

\end{document}